\documentclass[11pt,a4paper]{article}
\pdfoutput=1   % required for arXiv submission if pdflatex is used
\usepackage[utf8]{inputenc}
\usepackage{tabularx}
\usepackage{mathtools}
\usepackage{longtable}
\usepackage{times}
\usepackage{subcaption} 
\usepackage{enumerate}

\usepackage{amsmath}
\numberwithin{equation}{section}

\usepackage{alpha}
\hypersetup{%
   pdftitle    = {Ratio of flavour non-singlet and singlet
   	scalar density renormalisation parameters in Nf=3 QCD with Wilson quarks},
   pdfauthor   = {J. Heitger, F. Joswig, P. L. J. Petrak, and A. Vladikas},
}%
\usepackage{microtype}
\usepackage{lmodern}
\usepackage[outercaption]{sidecap}
\usepackage{macros}
\usepackage{xspace}
\usepackage{cancel}
\usepackage{array}% for extended column definitions 
\usepackage{siunitx}
\usepackage{lscape}
\usepackage{makecell}
\begin{document}
\interfootnotelinepenalty=10000
\renewcommand{\arraystretch}{1.3} % spreads out values in tables

\preprintno{%
MS-TP-21-02\\
\vfill
}

\title{%
Ratio of flavour non-singlet and singlet
scalar density renormalisation parameters in $N_\mathrm{f}=3$ QCD\\
with Wilson quarks}

\collaboration{\includegraphics[width=2.8cm]{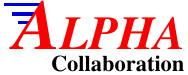}}

\author[ms]{Jochen Heitger}
\author[ms]{Fabian Joswig}
\author[ms]{Pia~L.~J.~Petrak}
\author[infn]{Anastassios Vladikas}
\address[ms]{Westf\"alische Wilhelms-Universit\"at M\"unster, Institut f\"ur Theoretische Physik,\\
Wilhelm-Klemm-Stra{\ss}e 9, 48149 M\"unster, Germany}
\address[infn]{INFN, ``Rome Tor Vergata'' Division, c/o Dipartimento di Fisica, \\ Via della Ricerca Scientifica~1, 00133 Rome, Italy}

\begin{abstract}
We determine non-perturbatively the normalisation factor $\rmsea \equiv \zs/\zss$, where $\zs$ and $\zss$ are the renormalisation parameters of the flavour non-singlet and singlet scalar densities, respectively. This quantity is required in the computation of quark masses with Wilson fermions and for instance the renormalisation of nucleon matrix elements of scalar densities. Our calculation involves simulations of finite-volume lattice QCD with the tree-level Symanzik-improved gauge action, $\nf = 3$ mass-degenerate $\rmO(a)$ improved Wilson fermions and Schr\"odinger functional boundary conditions. The slope of the current quark mass, as a function of the subtracted Wilson quark mass is extracted both in a unitary setup (where nearly chiral valence and sea quark masses are degenerate) and in a non-unitary setup (where all valence flavours are chiral and the sea quark masses are small). These slopes are then combined with $Z \equiv \zp/(\zs\za)$ in order to obtain $\rmsea$.  A novel chiral Ward identity is employed for the calculation of the normalisation factor $Z$. Our results cover the range of gauge couplings corresponding to lattice spacings below $0.1\,$fm, for which $\nf = 2+1$ QCD simulations in large volumes with the same lattice action are typically performed.
\end{abstract}

\begin{keyword}
Lattice QCD \sep
quark mass renormalisation \sep
Ward identities \sep
Schr\"odinger functional \sep
chiral Symmetry restoration with Wilson fermions%
\PACS{% 
11.15.Ha\sep %Lattice Gauge Theory
12.38.Gc\sep %Lattice QCD calculations
12.38.Aw     %general properties of QCD
}
\end{keyword}

\maketitle

%\tableofcontents
%\clearpage

% place here, otherwise minor issues with title page appear until fixed
\makeatletter
\g@addto@macro\bfseries{\boldmath}
\makeatother

\section{Introduction}
Scalar and pseudoscalar flavour singlet and non-singlet dimension-3 bilinear operators have the same anomalous dimension, since they belong to the same chiral multiplet. The same is true for their renormalisation parameters, provided  that the regularisation does not break chiral symmetry. Otherwise, the renormalisation parameters of the chiral multiplet components differ by finite terms. This is the case for the lattice regularisation with Wilson fermions. For example, the renormalisation parameters of the non-singlet scalar and pseudoscalar densities (denoted as $\zs$ and $\zp$, respectively) have a finite ratio which is a polynomial of the bare gauge coupling $g_0$. This ratio can be determined by chiral Ward identities;\footnote{In practice, distinct chiral Ward identities are used for the computation of the ratio $\zs/(\zp \za)$ and $\za$; the two results are subsequently multiplied to give $\zs/\zp$.}$\!$ see Refs.~\cite{Bochicchio:1985xa,Maiani:1987by}. Since $\zp$ and $\zs$ are scale dependent, imposing a renormalisation scheme is necessary to fix one of them, and the other can be obtained using the scheme independent ratio $\zs/\zp$.\footnote{Examples of renormalisation schemes are $\msbar$, RI/(S)MOM~\cite{Martinelli:1994ty,Sturm:2009kb}, the Schr\"odinger functional (SF)~\cite{Capitani:1998mq} and the chirally rotated Schr\"odinger functional ($\chi$SF)~\cite{Brida:2016rmy}.} In this way the renormalised scalar and pseudoscalar densities are defined consistently in the same scheme, with the same anomalous dimension and renormalisation group (RG) running, and chiral symmetry is restored in the continuum limit. The ratio $\zs/\zp$ has been computed for several gauge and Wilson fermion actions (standard, improved etc.) in the quenched approximation~\cite{Maiani:1987by,deDivitiis:1997ka,Bhattacharya:1999uq,Bhattacharya:2000pn,Guagnelli:2000jw,Bhattacharya:2005ss}, with two dynamical quarks ($\nf = 2$ QCD)~\cite{Fritzsch:2010aw}, and with three dynamical quarks ($\nf = 3$ QCD)~\cite{Bali:2016umi,deDivitiis:2019xla,Heitger:2020mkp,Bali:2020lwx}.

Far less progress has been made on the computation of the ratio of the renormalisation parameters of the non-singlet and singlet scalar densities, $\rmsea \equiv \zs/\zss$. For chirally symmetric regularisations  $\rmsea =1$ holds, while for Wilson fermions $\rmsea$ is a (finite) polynomial of the gauge coupling, arising from the sea fermion loops of the quark propagator. In the quenched approximation, $\rmsea = 1$. As explained in Ref.~\cite{Bhattacharya:2005rb}, the lowest-order non-trivial perturbative contribution to this quantity is a two-loop effect; i.e., $\rmsea = 1 + \rmO(g_0^4)$. In Ref.~\cite{Constantinou:2016ieh} the $\rmO(g_0^4)$ perturbative term has been calculated for several lattice actions. Non-perturbative estimates of this quantity have been reported in Ref.~\cite{Bali:2016umi} at two values of the gauge coupling for $\nf=2+1$ QCD with the tree-level Symanzik-improved gauge action~\cite{Luscher:1984xn} and the non-perturbatively improved Wilson-clover fermion action~\cite{Sheikholeslami:1985ij}. This is the regularisation chosen by the \textit{CLS} (\textit{Coordinated Lattice Simulations}) initiative which carries out QCD simulations with  $\nf = 2 + 1$ flavours, on large physical volumes, for a range of bare couplings corresponding to a hadronic regime~\cite{Bruno:2014jqa,Bruno:2016plf,Bali:2016umi,Mohler:2017wnb}. These \textit{CLS} ensembles are suitable for the computation of correlation functions, from which low-energy hadronic quantities can be evaluated. In parallel, our group is performing $\nf=3$ simulations in the same range of bare gauge couplings, but for small-volume lattices with Schr\"odinger functional boundary conditions and nearly-chiral quark masses. These ensembles are used for the numerical determination of the necessary renormalisation parameters and Symanzik improvement coefficients, see Refs.~\cite{Heitger:2017njs,Heitger:2020mkp,Heitger:2020zaq,Bulava:2016ktf,Bulava:2015bxa,Chimirri:2019xsv,deDivitiis:2019xla} that have various applications in lattice QCD when using this discretisation of Wilson fermions. The present work provides high-precision estimates of $\rmsea$ obtained in the same computational framework.

As seen from eq.~(\ref{eq:mren-mq}) below, $(\rmsea-1)$ contributes an $\rmO(g_0^4)$ term to the renormalisation of the quark masses~\cite{Bhattacharya:2005rb}. This is expected to be a small effect. Symanzik $\mathrm{O}(a)$ counterterms containing $\rmsea$ are often neglected in light quark mass determinations; cf.\ Ref.~\cite{Bruno:2019vup}. In practical computations, however, $r_\mathrm{m}$ can be relevant at $\mathrm{O}(a)$, especially when dealing with heavy flavours, and should be taken into account in order to achieve full $\mathrm{O}(a)$ improvement; see, for example, eq.~(2.13) in Ref.~\cite{Heitger:2021apz}.
Another application where $\rmsea$ plays a prominent r\^ole is the nucleon sigma-term, which is defined in terms of nucleon matrix elements of flavour singlet scalar densities; see Refs.~\cite{Bali:2011ks,Bali:2016lvx} for example and \cite{Aoki:2019cca,Ottnad:2020qbw,Green:2018vxw} for more recent works. A direct determination of $\zss$ is not as straightforward as that of $\zs$, the former also requiring the computation of two-boundary (``disconnected'') quark diagrams. This problem is circumvented by extracting $\zss$ as the product of $\zs$ and $\rmsea$.

Our computation of $\rmsea$ is based on the relation between the current (PCAC) mass $m$ and the subtracted quark mass $\mq$. Close to the chiral limit, $m(\mq)$ is a linear function with a slope that depends on the details of the QCD model being simulated. In a unitary theory with degenerate sea and valence quark masses, the slope of $m(\mq)$ is $Z \rmsea$, where $Z \equiv \zp/(\zs\za)$ and $\za$ is the non-singlet axial current normalisation. On the other hand, in a non-unitary theory with chiral valence subtracted quark masses ($\mq^{\rm val} = 0$) and small degenerate sea quark masses $\mq^{\rm sea} \neq 0$, the slope of $m(\mq^{\rm sea})$ is $Z(\rmsea-1)$. The two slopes are accessible from two distinct sets of measurements at several common values of the bare coupling $g_0$. The results are combined to give estimates of $\rmsea(g_0^2)$. This approach is described in Section~\ref{sec:masses-rm-Z}.

Alternatively, each of the two slopes $Z \rmsea$ and $Z(\rmsea-1)$ may be combined with an independent estimate of $Z$, such as the results of Refs.~\cite{deDivitiis:2019xla,Heitger:2020mkp}. In the present work we prefer to use a novel determination of $Z$, relying on a chiral Ward identity which differs from the one of Ref.~\cite{Heitger:2020mkp}. This identity is derived in Section~\ref{sec:ward_identity}.

In Section~\ref{sec:setup} we present our simulation setup for $\nf =3$ QCD with lattices of small physical volumes and Schr\"odinger functional boundary conditions; these serve to numerically implement the strategies outlined in the foregoing section. Most of our gauge field ensembles were already generated in the context of previous works; cf. Refs.~\cite{Heitger:2020mkp,Heitger:2020zaq,Bulava:2016ktf,Bulava:2015bxa,Chimirri:2019xsv,deDivitiis:2019xla}. Some new ensembles have also been generated, in order to cover the region close to the origin of the function $m(\mq)$ more evenly and asses its slope reliably.

Our results for $\rmsea$, based on various combinations of $Z \rmsea$,  $Z(\rmsea-1)$, and $Z$ are discussed in Section~\ref{sec:results}. Different determinations of $\rmsea$ are compared, allowing us to settle for a conservative final estimate with reliable systematic errors. Our final result is that of eq.~(\ref{eq:fit}). In Table~\ref{tab:rm_cls} we also list $\rmsea(g_0^2)$ for the $g_0^2$-values at which \textit{CLS} simulations are being performed for the computations of hadronic quantities in $\nf=2+1$ QCD.

In the final section we sum up our results and their uses in lattice QCD.
More detailed calculations and definitions of the correlation functions employed can be found in 
Appendix~\ref{app:sfcf} and \ref{app:corr-funct}. Comparison of $Z$ determinations and corresponding scaling tests can be found in Appendix~\ref{app:z}.

\section{Wilson quark masses}
\label{sec:masses-rm-Z}

In this section we recapitulate the basic quark mass definitions, namely subtracted and
current (PCAC) quark masses, and discuss how to obtain the products $Zr_\mathrm{m}$ and $Z(r_\mathrm{m}-1)$ from relations between the two. For any unexplained notation we refer to
Ref.~\cite{deDivitiis:2019xla}.
The starting point is the subtracted bare quark mass of flavour $i=1, \ldots , \Nf$,
\begin{equation}
\mqi \equiv m_{0,i} - m_{\rm crit} = \dfrac{1}{2a} \Big (\dfrac{1}{\hop_i}-\dfrac{1}{\hopcr} \Big )  \, ,
\label{eq:bare-mass}
\end{equation}
where $\hop_i$ is the hopping parameter for flavour $i$, $\hopcr$ its value in the chiral limit,
and $a$ is the lattice spacing. In terms of the subtracted masses $\mqi$, the corresponding renormalised 
quark masses are given by
\begin{eqnarray}   
m_{i,\rm R} = \zm  \Bigg [ \mqi \, + \, (\rmsea - 1) \dfrac{\Tr \Mq}{\Nf} \Bigg ] + \rmO(a)\,,
\label{eq:mren-mq}
\end{eqnarray}   
where $\Mq = {\rm diag}(m_{{\rm q},1}, \ldots , m_{{\rm q},\Nf})$ is the $\Nf \times \Nf$  bare quark mass matrix. 

We recall in passing that the renormalisation parameter $\zm(g_0^2,a\mu)$ depends on the renormalisation
scale $\mu$ and diverges logarithmically in the ultraviolet. It is the inverse of $\zs(g_0^2,a\mu)$, the renormalisation parameter of the flavour non-singlet scalar density operator.
A mass independent renormalisation scheme is implied throughout this work.
In such a scheme operator renormalisation parameters (e.g. $\zp, \zm, \zs$), current normalisations (i.e. $\za,\zv$) and
$\rmsea$ are functions of the squared bare gauge coupling $g_0^2$. In a non-perturbative determination at non-zero quark mass, they are affected by $\rmO(a\mqi)$, $\rmO(a \Tr\Mq)$, and $\rmO(a \lQCD)$ discretisation effects, which are part of their operational definition. As pointed out in Ref.~\cite{Bhattacharya:2005rb}, the term
$(\rmsea-1)$ multiplies $\Tr \Mq$, so it arises from a mass insertion in a quark loop. In perturbation theory it is a two-loop effect, contributing at $\rmO(g_0^4)$. Its non-perturbative determination is the main purpose of this paper. An important consequence of eq.~(\ref{eq:mren-mq}) is that a renormalised mass $m_{i,\rm R}$ goes to the chiral limit only when {\it all} subtracted masses $m_{{\rm q},1}, \dots, m_{{\rm q},\nf}$ vanish.

Alternatively, a bare current (PCAC) quark mass $m_{ij}$ can be {\it defined} through the following relation:
\begin{equation}
(\sdrv\mu)_x\evalbig{(\aimpr)^{ij}_\mu(x)\,\mathcal{O}^{ji}} = 2m_{ij} {\evalbig{{P}^{ij}(x)\,\mathcal{O}^{ji}}} \,.
\label{eqn:PCAC-mass}
\end{equation}
The quantity $m_{ij}$ is distinct from the subtracted bare quark masses, but it is related to the mass average $(\mqi + \mqj)/2$; see eq.~(\ref{eq:mPCAC-msub}) below.
The flavour non-singlet bare axial current and the pseudoscalar density are given by
\begin{equation}
\label{eq:PCACgen}
A_{\mu}^{ij}(x)  \equiv  \psibar_i(x)\,\dirac\mu\dirac5\,\psi_j(x)\,, \qquad 
P^{ij}(x) \equiv  \psibar_i(x)\,\dirac5\,\psi_j(x)  \,,
\end{equation}
with indices $i,j$ denoting two distinct flavours ($i \neq j$). The pseudoscalar density $P^{ij}$ and the current $(\aimpr)^{ij}_\mu \equiv A^{ij}_\mu+a\ca\sdrv\mu P^{ij}$ are Symanzik-improved in the chiral limit, with the improvement coefficient $\ca(g_0^2)$ being in principle only a function of the gauge coupling.
In these definitions, $\sdrv\mu$~denotes the average of the usual forward and backward derivatives.\footnote{The forward derivative is defined as $a\partial_\mu f(x) \equiv f(x+a\hat\mu) - f(x)$ and the backward derivative as $a\partial_\mu^\ast f(x) \equiv f(x) - f(x-a\hat\mu)$.} The source operator $\mathcal{O}^{ji}$ is defined in a region of space-time that does not include the point $x$, so as to avoid contact terms.
In the $\rmO(a)$ improved theory, the renormalised axial current and pseudoscalar density are
\begin{align}
(\ar)^{ij}(x) &= \za(\gosq) (A_{\rm I})_{\mu}^{ij}(x) +  \rmO(am_\mathrm{q},a^2)\,,\\
(\pr)^{ij}(x) &= \zp(\gosq,a\mu) P^{ij}(x) + \rmO(am_\mathrm{q},a^2)\,.
\label{eqn:ren-off-diag-P}
\end{align}
The normalisation of the axial current $\za(g_0^2)$ is scale independent, depending only on the squared gauge coupling $g_0^2$. The renormalisation parameter $\zp(g_0^2,a\mu)$ (determined, say in the Schr\"odinger functional scheme of Ref.~\cite{Capitani:1998mq})
additionally depends on the renormalisation scale $\mu$ and diverges logarithmically in the ultraviolet.  The PCAC relation, expressed by renormalised fields,
\begin{align}
(\sdrv\mu)_x \evalbig{\, (\ar)^{ij}_\mu(x)\ \mathcal{O}^{ji}\, } &=  
(\mri{i}+\mri{j})\,\evalbig{\, (P_\mathrm{R})^{ij}(x)\ \mathcal{O}^{ji}\,} \,,
\label{eqn:renorm-PCAC-relation-pm}
\end{align}
valid up to discretisation effects in the continuum, combined with eqs.~(\ref{eqn:PCAC-mass})--(\ref{eqn:ren-off-diag-P}), implies that
\begin{equation}
\label{eq:renmassPCAC}
\dfrac{m_{i,\rm R} + m_{j,\rm R}}{2} =  \dfrac{\za}{\zp} \m_{ij}  + \rmO(am_\mathrm{q},a^2)\,.
\end{equation}
If we calculate the average mass $(m_{i\rm R} + m_{j\rm R})/2$ from eq.~(\ref{eq:mren-mq}) and equate the result to the r.h.s of eq.~(\ref{eq:renmassPCAC}), we obtain an expression which relates subtracted and PCAC bare masses:
\begin{equation}
\label{eq:mPCAC-msub}
m_{ij} =  Z  \Bigg [ \dfrac{(\mqi + \mqj)}{2} +  (\rmsea - 1) \dfrac{\Tr \Mq}{\NF}  \Bigg ]  +   \rmO(am_\mathrm{q},a^2)\,,
\end{equation} 
where  the product of the renormalisation parameters $Z(g_0^2) \equiv \zp(g_0^2,\mu)/(\zs(g_0^2,\mu)\za(g_0^2))$ is scale independent.
We now exploit eq.~(\ref{eq:mPCAC-msub}) in two ways:

({\bf 1}) In a theory with mass-degenerate quarks ($\mqi = \mqj = \Tr \Mq/\NF$), it reduces to 
\begin{align}
m &= Z \rmsea \mq + \rmO(am_\mathrm{q},a^2) \\
&= Z \rmsea \dfrac{1}{2a} \Big (\dfrac{1}{\hop}-\dfrac{1}{\hopcr} \Big ) +  \rmO(am_\mathrm{q},a^2)\,.\label{eq:mPCAC-msub-2}
\end{align} 
In the above equation, flavour indices have been dropped from the quark masses $m_{ij}, \mqi$ and the hopping parameter $\hop_i$. This simplification of notation will be adopted on most occasions below.
Thus, modelling the current quark mass $a m$ as a function of $1/\hop$ for values of $\hop$ close to $\hopcr$, we obtain the latter as the root of the function $am(1/\hop)$ and the combination $Z \rmsea$  as the slope of the same curve.

({\bf 2}) Once the critical hopping parameter  $\hopcr$ is available from the previous step ({\bf 1}), we use a non-unitary setup where valence and sea quarks of the same flavour have different bare subtracted masses $\mqi^{\rm val} \neq \mqi^{\rm sea}$. In eq.~(\ref{eq:mPCAC-msub}), masses $\mqi$ and $ \mqj$ on the r.h.s. are valence quark contributions, while $\Tr \Mq$ stands for the trace of sea quark masses; see Refs.~\cite{Bhattacharya:2005rb,deDivitiis:2019xla} for detailed explanations. In particular, we set $\hopval = \hopcr$, so as to ensure $\mq^{\rm val}=0$ for all valence flavours. Moreover sea quark masses are taken to be small, degenerate, and non-zero (i.e. $\hopsea \neq \hopcr$, ensuring $\mq^{\rm sea} \neq 0$ for all sea flavours). With these conditions, the current quark mass of eq.~(\ref{eq:mPCAC-msub}) reduces to 
\begin{equation}
\label{eq:mPCAC-msub-3}
m =  Z  (\rmsea - 1) \mq^{\rm sea}  +   \rmO(am_\mathrm{q},a^2) \, .
\end{equation} 
It is remarkable that with  non-zero bare subtracted sea quark masses (i.e. $\mq^{\rm sea} \neq 0$), all current quark masses in this setup are not chiral (i.e. $m_{ij} \neq 0, \forall i,j$), even if all subtracted valence quark masses vanish (i.e. $\mqi^{\rm val} = 0, \forall i$). From eq.~(\ref{eq:mPCAC-msub-3}) we see that, if we compute $a m$ as a function of $a \mq^{\rm sea}$ for several sea masses, the slope of the functions gives an estimate of $Z (\rmsea - 1)$.

The two slopes $Z \rmsea$ and $Z (\rmsea - 1)$, computed in the two different settings described above, but at the same gauge couplings $g_0^2$, can be combined yielding estimates of $\rmsea(g_0^2)$; see Subsection~\ref{sec:rm-results} for details.
We stress that the above discussion concerns relations which suffer from $\rmO(a)$ discretisation effects. For the quark masses, such effects may be removed by introducing Symanzik counterterms, leaving us with $\rmO(a^2)$ discretisation errors. These counterterms have been worked out in Refs.~\cite{Luscher:1996sc,Bhattacharya:2005rb}. In Ref~\cite{Bhattacharya:2005rb} (see also eq.~(2.10) of Ref.~\cite{deDivitiis:2019xla}) the full $\rmO(a\mq)$ contributions, omitted in eq.~(\ref{eq:mPCAC-msub}) above, are written down explicitly. Such contributions are complicated and taking them all into account could compromise the numerical stability of our procedure to extract the quantities in question. We prefer a simpler and more robust strategy, consisting of working with small quark masses so that $\rmO(a)$-effects in eq.~(\ref{eq:mPCAC-msub}) may be safely dropped. This must of course be checked {\it a posteriori}, by ensuring that the function $m(\mq)$ is linear close to the origin, where our simulations are performed.
The only improvement coefficients used in this work are $c_\mathrm{sw}$ of the clover action and $\ca$, of the axial current (entering the PCAC mass).

There is an important subtlety concerning results obtained with Wilson fermions in a Symanzik-improved setup: the bare parameters of the theory (i.e. the gauge coupling $g_0^2$ and the $\Nf = 2 + 1$ quark masses) are to be varied, while staying on lines of constant physics within systematic uncertainties of $\rmO(a^2)$. In particular, if the improved bare gauge coupling~\cite{Luscher:1996sc}
\begin{equation}
\label{eq:g0-tilde}
\tilde g_0^2 \equiv \gosq \Big ( 1 + \dfrac{1}{\Nf} b_g(\gosq) a  \Tr \Mq \Big )
\end{equation} 
is kept fixed in the simulations, so is the lattice spacing, with fluctuations being attributed to  $\rmO(a^2)$ effects~\cite{Bruno:2014jqa}. This implies that, once $\kappa^\mathrm{crit}$ has been evaluated as a function of $g_0^2$, (re)normalisation parameters and
improvement coefficients should be treated as functions of $\tilde g_0^2$, rather than $\gosq$; e.g. $\za(\tilde g_0^2), \zp(\tilde g_0^2,a\mu), Z(\tilde g_0^2), \rmsea(\tilde g_0^2)$ etc. To the extent that we are working in the chiral limit, or very close to it (i.e. very light quark masses), this difference is immaterial. This is why in the present work we always express our results as functions of $\gosq$. However, when they are to be used away from the chiral limit at low-energy scales
(see Refs.~\cite{Bruno:2016plf,Bruno:2019vup}), this difference must be taken into account properly. We shall elaborate further on this point when summarising our work in Section~\ref{sec:conclusions}.
\section{\texorpdfstring{Ward identity determination of $Z$}{Ward identity determination of Z}}
\label{sec:ward_identity}
In the previous section we have shown how the quantities $Z \rmsea$ and $Z(\rmsea - 1)$ can be estimated from relations between suitably chosen
current and subtracted Wilson quark masses. They may then straightforwardly be combined to give $\rmsea$ and $Z$.
The latter quantity has already been measured in our setup ($\NF=3$ lattice QCD with Schr\"odinger functional boundary conditions) in two ways: either by using appropriate combinations of current and subtracted quark masses with different flavours~\cite{deDivitiis:2019xla}, or from chiral Ward identities~\cite{Heitger:2020mkp} via $Z \equiv \zp/(\za \zs)$. Here we will describe yet another direct method, based on a new Ward identity, very similar to the one of Ref.~\cite{Heitger:2020mkp}. The reader is referred to that work for details, notation etc.

We consider a product of two composite operators $\cO \equiv S^b(y) \cO^c$, defined as
\begin{eqnarray}
S^b(y) &\equiv&  \mathrm{i} \bar \psi(y) T^b \psi(y)  \nonumber \\
\cO^c &\equiv&  \mathrm{i} \dfrac{a^6}{L^3} \sum_{\bf u,v} \bar \zeta({\bf u}) \gamma_5 T^c  \zeta({\bf v})\,,
\end{eqnarray}
where $T^b$ and  $T^c$ are generators of $SU(N_\mathrm{f})$.
The former operator is the flavour non-singlet scalar density, located in the bulk of space-time, while the latter resides at the $x_0 = 0$ Dirichlet time boundary of the Schr\"odinger functional.\footnote{For reasons of convenience, we have adopted a slightly different notation in this section: the flavour content of operators like $S^b$ or $\cO^c$ is determined by a single flavour index $b$ or $c$, corresponding to its flavour matrix $T^b$ or $T^c$.  The fermion fields of these operators $\psi$ and $\bar \psi$ are columns in flavour space. This is to be contrasted to the notation of Section~\ref{sec:masses-rm-Z}, where we have introduced operators like $P^{ij}$ and $\mathcal{O}^{ji}$, which have explicit indices, referring to the flavour of fields $\psi_j, \bar \psi_i$ etc.} The Ward identity of interest is obtained by performing axial variations on $\cO$ in a region $R$, chosen to be the space-time volume between the hyper-planes at $t_1$ and $t_2$ where $t_1<t_2$. With $\cO^c$ lying outside $R$, we have $\delta_{\rm A} \cO = [\delta_{\rm A} S^b(y)]\cO^c$ and
\begin{equation}
\delta_{\rm A} S^b(x) =  \epsilon^a \Big [ d^{abe} P^e(x) + \dfrac{\delta^{ab}}{\NF} \bar \psi(x) \psi(x) \Big ] \, .
\end{equation}
In what follows we simplify matters by always working with $a \neq b$, so as to eliminate the second contribution on the r.h.s. of the above expression. In analogy to the derivation exposed in Ref.~\cite{Heitger:2020mkp}, we arrive at the formal continuum Ward identity
\begin{eqnarray}
&& \int \mathrm{d}^3{\bf y} \int \mathrm{d}^3{\bf x} \Big \langle \Big [A_0^a(t_2;{\bf x}) - A_0^a(t_1;{\bf x}) \Big ] S^b(y_0;{\bf y}) \cO^c \Big \rangle \nonumber \\
&& - 2m \int \mathrm{d}^3{\bf y} \int \mathrm{d}^3{\bf x} \int_{t_1}^{t_2} \mathrm{d}x_0 \langle P^a(x_0;{\bf x}) S^b(y_0;{\bf y}) \cO^c \rangle
\label{eq:SPWImassint2} \\
&& = - d^{abe} \int \mathrm{d}^3{\bf y} \,\, \langle P^e(y) \cO^c \rangle \, .
\nonumber
\end{eqnarray}

Next we adapt the previous formal manipulations to the lattice regularisation with Schr\"odinger functional boundary conditions. The pseudoscalar operator $\cO^c$ is defined on the $x_0=0$ time boundary. Ward identity (\ref{eq:SPWImassint2}) then becomes:
\begin{eqnarray}
&&  \za \zs  a^6 \Bigg \{ \sum_{{\bf x},{\bf y}} \, \left\langle \Big [ (A_{\rm I})^a_0(t_2; {\bf x}) - (A_{\rm I})^a_0(t_1; {\bf x}) \Big ]\, S^b(y_0;{\bf y}) \,  \cO^c \right\rangle
\nonumber \\
&& -  2 a m \sum_{{\bf x},{\bf y}} \sum_{x_0 = t_1}^{t_2} w(x_0) \, \langle P^a(x_0;{\bf x}) \, S^b(y_0;{\bf y}) \,  \cO^c \rangle \Bigg \}
 \label{eq:SPImassintlatt}\\
&& =  - d^{abe} \zp \,\,  a^3 \sum_{\bf y} \langle  \, P^e(y) \, \cO^c \rangle  +  \rmO(am,a^2) \,.
\nonumber
\end{eqnarray}
In this expression repeated flavour indices $e$ are summed, as usual. The weight factor is $w(x_0) = 1/2$ for $x_0 \in \{t_1,t_2\}$  and $w(x_0) = 1$ otherwise. It is introduced in order to implement the trapezoidal rule for discretising integrals. Quark masses are degenerate and $m$ is the current quark mass.

\begin{figure}[t]
	\centering
	\begin{subfigure}[b]{0.32\textwidth}
		\centering
		\includegraphics[width=\textwidth]{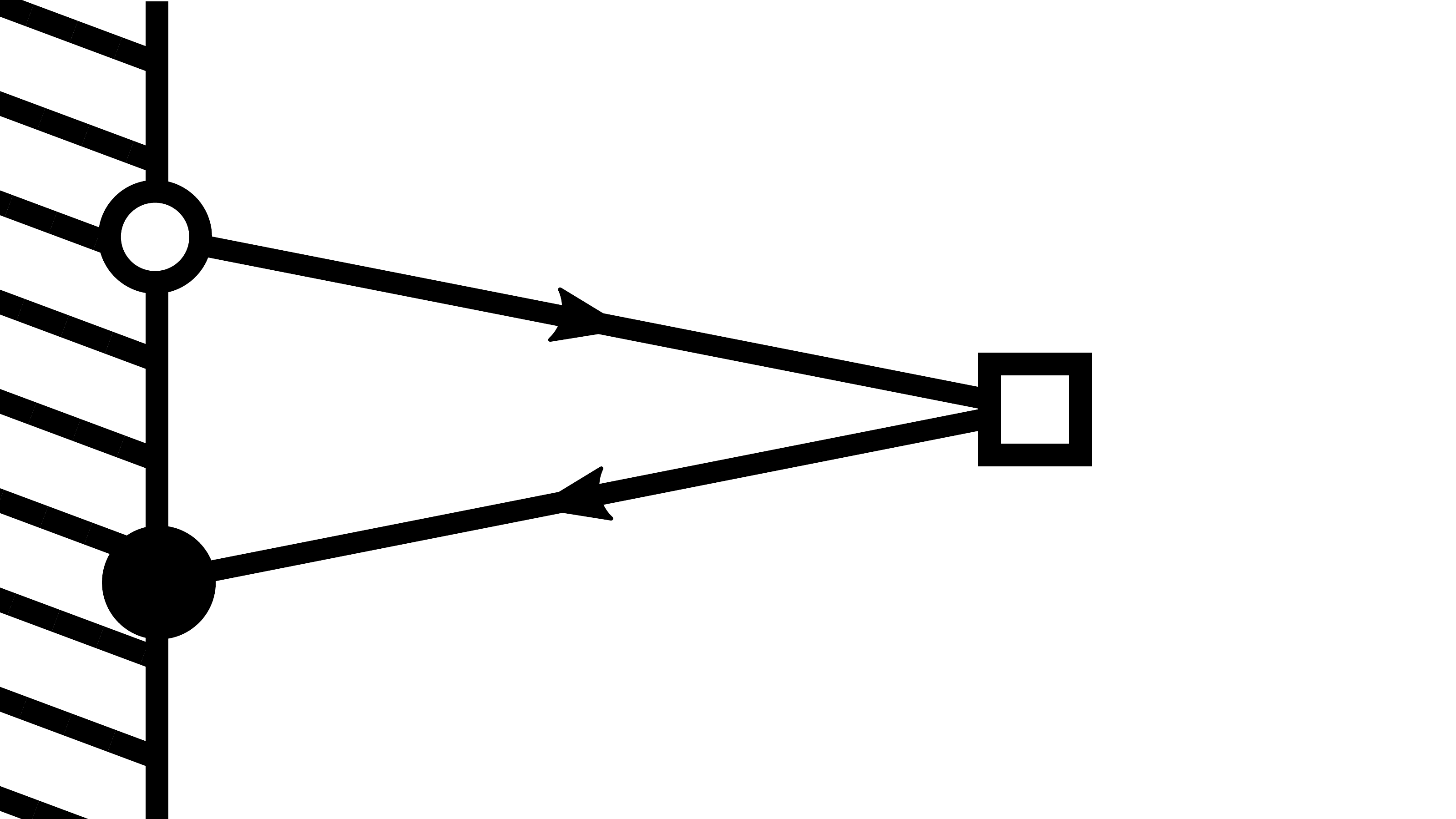}
		\caption{Diagram $\fp$}
	\end{subfigure}
	\begin{subfigure}[b]{0.32\textwidth}
		\centering
		\includegraphics[width=\textwidth]{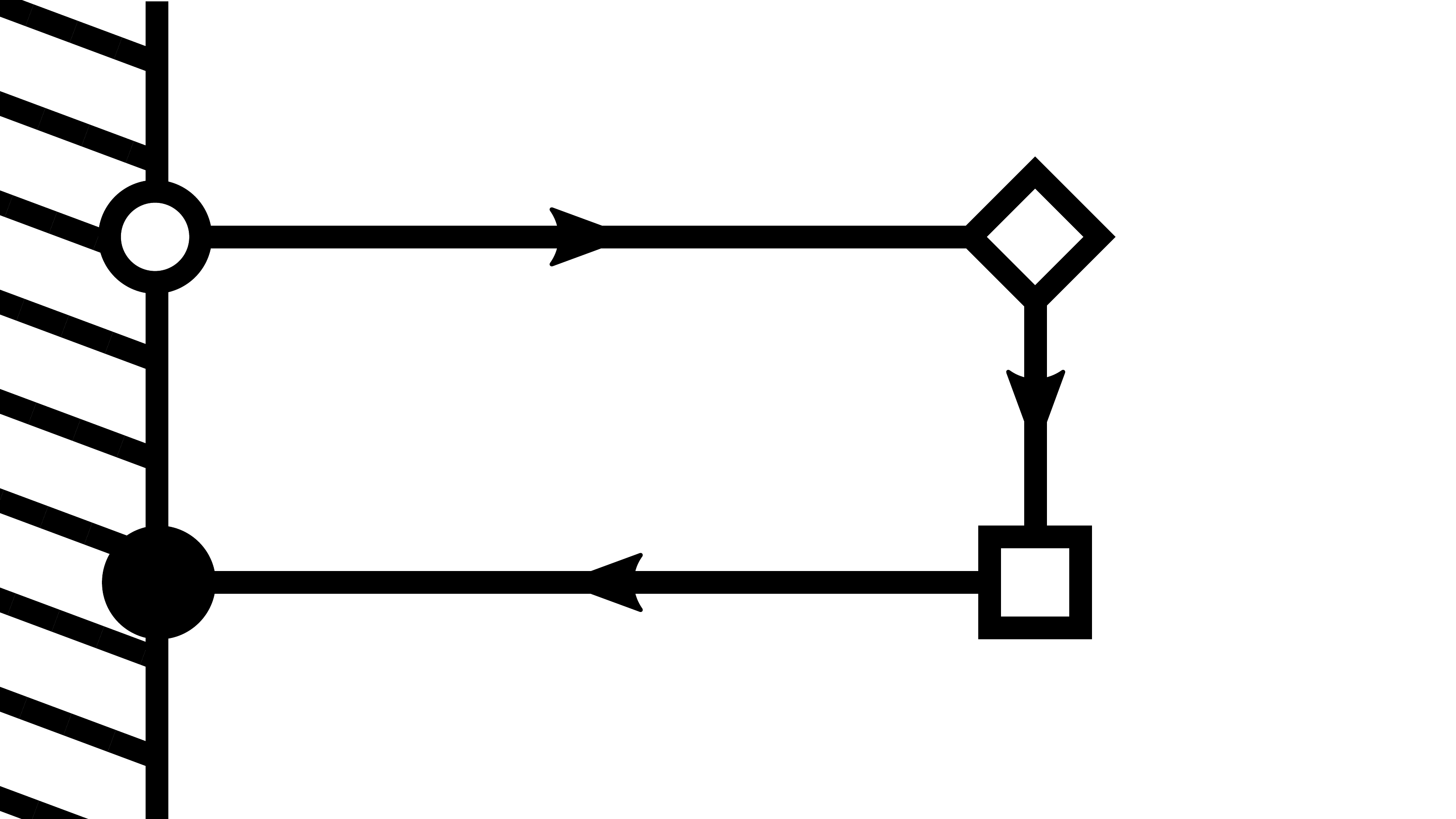}
		\caption{Diagram $f_{\rm A S;1}$}
	\end{subfigure}
	\begin{subfigure}[b]{0.32\textwidth}
		\centering
		\includegraphics[width=\textwidth]{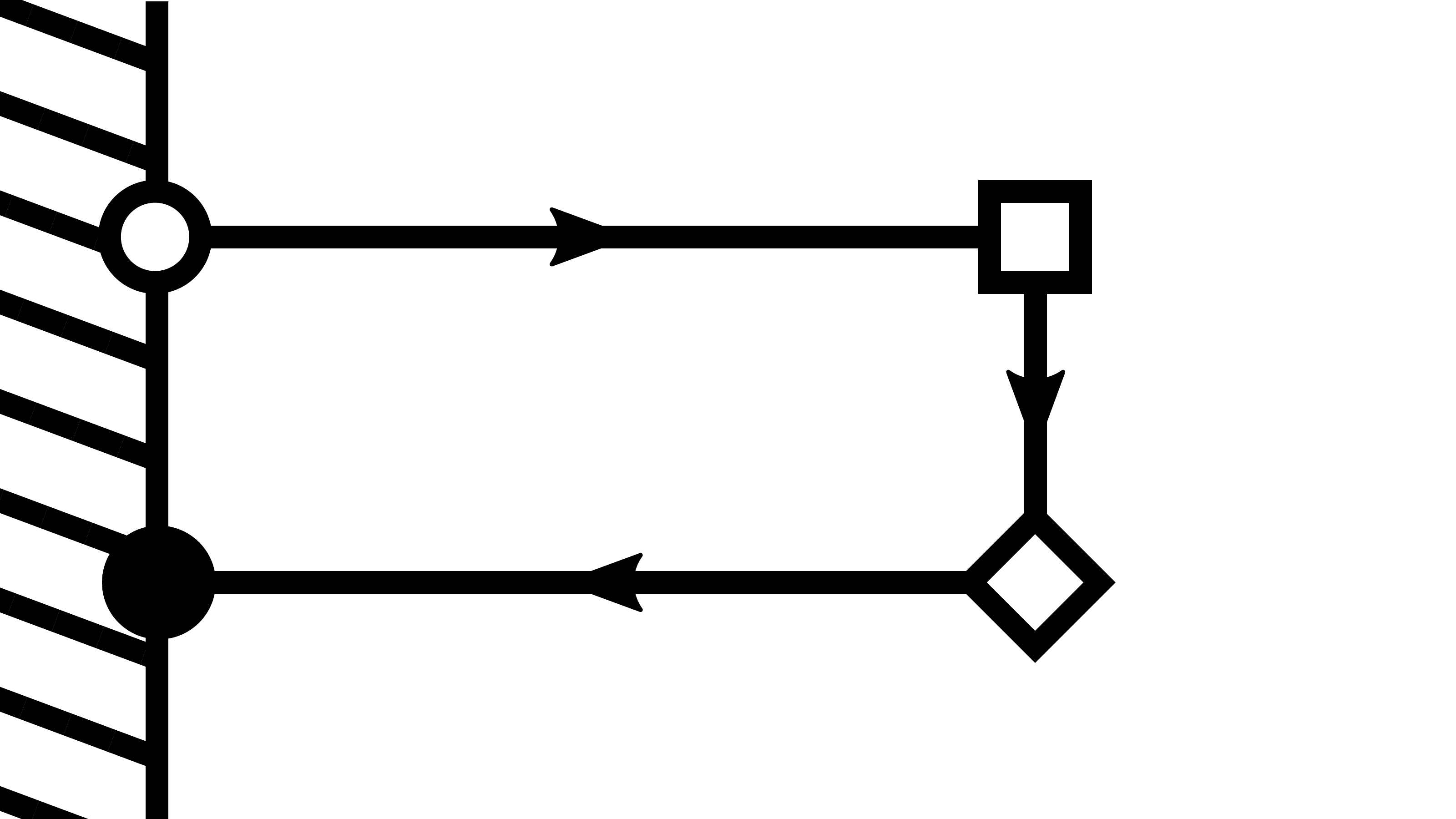}
		\caption{Diagram $f_{\rm A S;2}$}
	\end{subfigure}
	\caption{The trace diagrams contributing to the expectation values of $\fp$, defined in eq.~(\ref{eq:trBO}) (diagram (a)) and $\fas$, defined in eq.~(\ref{eq:trBOO-2}) (diagrams (b) and (c)). The wall represents the time slice $x_0 = 0$ with a $\gamma_5$ Dirac matrix between circles. The squares in the bulk represent either the insertions of a pseudoscalar operator $P(y)$  (diagram (a)) or a scalar operator $S(y)$ (diagrams (b) and (c)). The diamonds stand for an axial operator $A_0(x)$. The open circles correspond to the boundary fields $\zeta$, while the filled circles denote $\bar \zeta$. The diagrams schematically represent traces, formed by starting from any point and following the lines (quark propagators) until we close the loop. The time ordering of points $x$ and $y$ is left unspecified in these diagrams.}
	\label{fig:WI}
\end{figure}

The last step is to perform the Wick contractions in Ward identity~(\ref{eq:SPImassintlatt}). How this is done is explained in Appendix~\ref{app:corr-funct}; eventually, flavour factors drop out and we are left  with a Ward identity that translates into traces of products of quark propagators and $\gamma$-matrices, graphically depicted in Fig.~\ref{fig:WI}. Solving for $Z$ we get
\begin{eqnarray}
Z \equiv  \dfrac{\zp}{\za \zs} = {-} \dfrac{\fas^\mathrm{I}(t_2,y_0) -  \fas^\mathrm{I}(t_1,y_0) - 2 am \ftps(t_2,t_1,y_0)}{\fp(y_0)} + \rmO(am,a^2)\,,
\label{eq:WIf}
\end{eqnarray}
where dependencies are suppressed on the l.h.s.
Assuming that we work in the chiral limit (or with nearly-vanishing quark masses, so that $\rmO(am)$ effects may be safely neglected),
the above Ward identity is valid up to $\rmO(a^2)$ discretisation errors in lattice QCD with Wilson quarks. In this spirit, terms proportional to Symanzik $b$-coefficients may also be safely ignored.\footnote{This is even true for light (up/down, strange) non-chiral quark masses, as explicitly demonstrated in Ref.~\cite{Bruno:2019vup}, using the $b$-coefficients of Ref.~\cite{deDivitiis:2019xla}.}
The renormalisation factor of the external source $\cO^c$ is not taken into consideration, as it cancels out in the ratio (\ref{eq:WIf}). 
The term proportional to the current quark mass $m$ may also be dropped close to the chiral limit, but since we are working with masses which are not strictly zero, it could be advantageous to keep it in practice. In fact, it was found in Refs.~\cite{Bulava:2016ktf,Heitger:2020mkp} that this term stabilizes the chiral extrapolation leading to smaller errors. This turns out to be true also in our case, as we will show in Subsection~\ref{sec:z} and Fig.~\ref{z_plateau}.

It is interesting to compare Ward identity~(\ref{eq:SPWImassint2}) with those of Ref.~\cite{Heitger:2020mkp}:
\begin{itemize}
\item
In Ref.~\cite{Heitger:2020mkp} the flavour factors gave rise to a multitude of identities, which were combined in order to increase the signal-to-noise ratio, while here we only have one identity. On these grounds one could expect that the numerical results of Ref.~\cite{Heitger:2020mkp} are more precise than the ones from the Ward identity introduced here.
\item
On the other hand, the identities of Ref.~\cite{Heitger:2020mkp} involved: (i) correlation functions with one operator insertion in the bulk of the lattice and one wall source at each time slice; cf.\ Fig.~1 in that work; (ii) correlation functions with two operator insertions in the bulk and one wall source at each time slice; cf.\ Fig.~2 in that work. Here we have: (i) a correlation function with one operator insertion in the bulk and one wall source; (ii) correlation functions with two operator insertions in the bulk and one wall source. These somewhat simpler correlation functions illustrated in Fig.~\ref{fig:WI} above are expected to have less statistical fluctuations. From this point of view, the results of the present work are expected to gain in accuracy.
\end{itemize}
Thus, one of our aims is to establish which of the two approaches leads to more accurate results. This is discussed in Subsection~\ref{sec:z} and Appendix~\ref{app:z}.

\section{Numerical setup}
\label{sec:setup}
\begin{table}[th!]
	\centering
	\begin{tabular}{lllrrll}
		\toprule
		$(L/a)^3\times T/a$&$\beta$&$\kappa$&\#REP&\#MDU&ID&$a$ (in fm)\\
		\midrule
		$12^3\times17$&3.3&0.13652&20&20480&A1k1&$0.1045(18)$\\
		&&0.13648&5&6876
		&A1k3&\\
		&&0.13650&20&96640&A1k4&\\
		$12^3\times18$&&$0.13612$&4&41600&\textit{A3k1}&\\
		&&$0.13627$&4&41600&\textit{A3k2}&\\
		&&$0.13593$&4&41600&\textit{A3k3}&\\
		&&$0.136444$&4&41600&\textit{A3k4}&\\
		&&$0.136575$&4&41600&\textit{A3k5}&\\
		&&$0.136385$&4&41600&\textit{A3k6}&\\
		\midrule
		$14^3\times21$&3.414&0.13690&32&38400&E1k1&$0.08381(68)$\\
		&&0.13695&48&57600&E1k2&\\
		$14^3\times20$&&0.13656&18&60480&\textit{E2k1}&\\
		&&0.13675&18&60480&\textit{E2k2}&\\
		\midrule
		$16^3\times 23$&3.512&0.13700&2&20480&B1k1&$0.06954(43)$\\
		&&0.13703&1&8192&B1k2&\\
		&&0.13710&2&16384&B1k3&\\
		&&0.13714&1&27856&B1k4&\\
		$16^3\times 24$&&0.13677&1&25904&\textit{B3k1}&\\
		\midrule
		$20^3\times29$&3.676&0.13680&1&7848&C1k1&$0.05170(42)$\\
		&&0.13700&4&15232&C1k2&\\
		&&0.13719&4&15472&C1k3&\\
		\midrule
		$24^3\times35$&3.810&0.13711875582&5&8416&D1k1$^\ast$&$0.04175(70)$\\
		&&0.13701&2&6424&D1k2&\\
		&&0.137033&8&85008&D1k4&\\
		\bottomrule
	\end{tabular}
	\caption{Simulation parameters $L$, $T$, $\beta$, $\kappa$, the number of replica \#REP and the number of molecular dynamics units \#MDU for the ensembles labelled by ID. Ensembles highlighted in {\it italics} were newly generated for this study while the remaining ones were already used in previous investigations (see, for example Ref.~\cite{deDivitiis:2019xla}). The ensemble D1k1 marked by an asterisk is only used for the determination of the PCAC masses. The lattice spacings $a$ are obtained by interpolating the results of Ref.~\cite{Bruno:2016plf} with a polynomial fit. All configurations are separated by 8 MDU's except for the ensembles A1k3 (4 MDU's) and D1k4 (16 MDU's).}
	\label{tab:sim_table}
\end{table}

We employ the tree-level Symanzik-improved gauge action and $N_\mathrm{f}=3$ mass-degenerate $\mathrm{O}(a)$ improved Wilson fermions. For the corresponding improvement coefficient $c_\mathrm{sw}$ we use the non-perturbative determination of Ref.~\cite{Bulava:2013cta}. As already indicated, we impose Schr\"odinger functional boundary conditions at the temporal boundaries of the lattice.
The Schr\"odinger functional setup is highly suitable for massless renormalisation schemes, since nearly-vanishing quark masses are accessible in numerical calculations due to the spectral gap of the Dirac operator. This gap is imposed by the boundaries, so that the quark mass dependence can be mapped out reliably in the vicinity of the chiral point.
The generation of the gauge field configurations is performed with the {\ttfamily{openQCD}} code \cite{openqcd} which employs the RHMC algorithm \cite{Kennedy:1998cu,Clark:2006fx} for the third quark. 
 
All gauge field ensembles used in this study are summarized in Table~\ref{tab:sim_table} and lie on a line of constant physics (LCP), defined by a fixed spatial extent of $L \approx 1.2 \, \si{fm}$ and $T/L \approx 3/2$. 
%The tuning was based on the two-loop beta-function; see Ref.~\cite{Bulava:2015bxa}.
%This ensures that our estimates of $r_\mathrm{m}$ and $Z$ become smooth functions of the lattice spacing, with relevant higher-order ambiguities vanishing monotonically.
The tuning was guided by the two-loop beta-function; see Ref.~\cite{Bulava:2015bxa}.
Provided that this perturbative approximation is satisfactory in the case at hand,
this ensures that our estimates of $\rmsea$ and $Z$ become smooth
functions of the lattice spacing, with higher-order ambiguities vanishing
monotonically. In Ref.~\cite{Heitger:2020zaq} it was explicitly shown that $L$ 
is constant up to $\rmO(a)$ cut-off effects across the coupling range also considered
in the present work. We thus expect our final
results for $\rmsea$ and $Z$ to only be affected by $\rmO(a^2)$ effects.
These are beyond the order we are interested in and they are treated as an ambiguity
that extrapolates to zero in the continuum
limit.\footnote{More precisely, the results on scale setting for our lattice 
action from Ref.~\cite{Bruno:2016plf} have been used in Ref.~\cite{Heitger:2020zaq} 
in order to demonstrate numerically that the deviation from a constant value of $L$ in 
physical units is proportional to the lattice spacing $a$. As the latter work uses the
configuration ensembles and range of non-perturbative bare couplings used also in the present
paper, our simulation parameters define a LCP up to $\rmO(a)$
lattice artefacts, so that the discretisation effects of $\rmsea$ and $Z$ 
are $\rmO(a^2)$  in the $\rmO(a)$ improved
theory.}

The gauge ensembles highlighted in {\it italics} were newly generated for this study, while the remaining ones were already used in previous investigations; see Refs.~\cite{Heitger:2017njs,Heitger:2020mkp,Heitger:2020zaq,Bulava:2016ktf,Bulava:2015bxa,Chimirri:2019xsv,deDivitiis:2019xla}.\footnote{In a setup with heavy sea quarks and very light valence quarks we approach a quenched-like situation in which exceptional configurations are to be expected; cf. Ref.~\cite{Luscher:1996ug} where a similar situation is discussed. In a careful analysis we identified only one gauge field configuration in the ensemble E2k1, with an exceptionally small eigenvalue of the massless Dirac operator. This leads to very large values of the correlation functions $f_\mathrm{P}$ and $f_\mathrm{A}$. We have discarded this exceptional configuration.}
These additional ensembles allow for a more even and wider spread of bare quark masses around the chiral point for each value of $\beta$, which enables a more precise extraction of the slopes corresponding to $Zr_\mathrm{m}$ and $Z\left(r_\mathrm{m}-1\right)$ as explained in Section~\ref{sec:masses-rm-Z}.
Since a newer version of the {\ttfamily{openQCD}} code was utilised for the generation of the ensembles, the time extent $T/a$, which was odd in the pre-existing ensembles, is even for the new ones.
For all ensembles we use tree-level boundary $\mathrm{O}(a)$ improvement for both the gauge and fermion fields (i.e. the appropriate $\ct, \cttil$ values) as if the time extents were even. The fact that an odd time extent alters the tree-level value of $\ct$, depending on the definition of the line of constants physics \cite{PerezRubio:2011cr}, affects the current quark masses below the precision achieved here, 
%only at the level of negligible higher-order ambiguities, 
as explicitly demonstrated in Ref.~\cite{Bulava:2015bxa}.

All Schr\"odinger functional correlation functions required for our numerical investigations are $\mathrm{O}(a)$ improved. In this context we only require the improvement coefficient $c_\mathrm{A}$, non-perturbatively known from Ref.~\cite{Bulava:2015bxa}.
Since the Markov chain Monte Carlo sampling of the gauge field configurations suffers from critical slowing down of the topological charge for smaller lattice spacings (see Ref.~\cite{DelDebbio:2002xa}), we project our data to the trivial topological sector as suggested in Ref.~\cite{Fritzsch:2013yxa}, in order to account for the insufficient sampling of all topological sectors. 
For the analysis of the statistical errors we employ the $\Gamma$-method \cite{Wolff:2003sm}. We account for the remaining critical slowing down of the Monte Carlo algorithm by attaching a tail to the autocorrelation function, as suggested in Ref.~\cite{Schaefer:2010hu}. The corresponding slowest mode is estimated from the autocorrelation time of the boundary-to-boundary correlation function $F_1^{ij}$, defined in Appendix \ref{app:sfcf}.
The error analysis is carried out with a python implementation of the $\Gamma$-method, using automatic differentiation for the error propagation as proposed in Ref.~\cite{Ramos:2018vgu}.

\section{Analysis details and results}
\label{sec:results}
In the following we present our analysis which eventually leads to several estimates for the ratio of the renormalisation parameters of the non-singlet and singlet scalar densities, $\rmsea$. We will first describe how we obtain $Zr_\mathrm{m}$, $Z(r_\mathrm{m}-1)$, and $Z$ individually and then discuss several ways of combining the three into $r_\mathrm{m}$. As a final result we provide an interpolation formula for $r_\mathrm{m}$ and extract its value at the bare couplings of large-volume \textit{CLS} simulations \cite{Bruno:2014jqa,Bali:2016umi,Mohler:2017wnb}.

\subsection{Quark mass slopes}
\label{sec:quark-mass-slopes}
	
\begin{figure}[t]
	\centering
	\begin{subfigure}[b]{0.45\textwidth}
		\centering
		\includegraphics[width=1.1\textwidth]{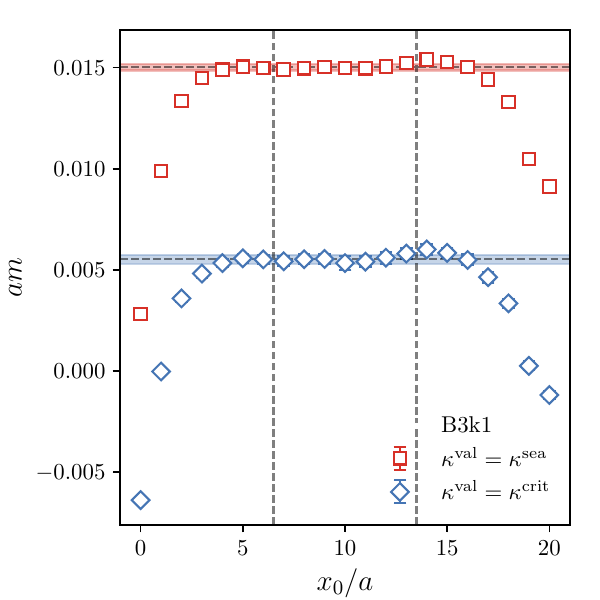}
	\end{subfigure}
	\hspace{0.5cm}
	\begin{subfigure}[b]{0.45\textwidth}
		\centering
		\includegraphics[width=1.1\textwidth]{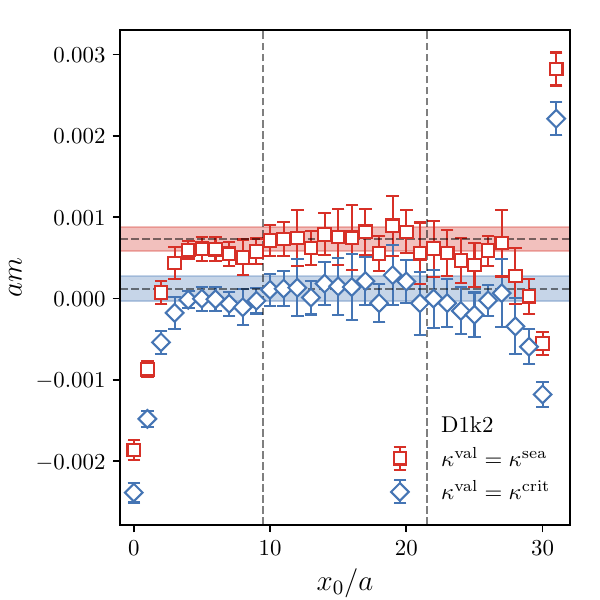}
	\end{subfigure}
	\caption{PCAC mass $am$ as a function of time $x_0/a$ for ensembles B3k1 (left) and D1k2 (right). Squares are results obtained in the unitary setup, while diamonds are results obtained in the non-unitary setup. The final estimate for $m$ is obtained by averaging results in the time interval $[T/3,2T/3]$, indicated by the dashed vertical lines.}
	\label{pcac_plateau}
\end{figure}

As described in Section~\ref{sec:masses-rm-Z}, the quantities $Zr_\mathrm{m}$ and $Z(r_\mathrm{m}-1)$ can be extracted from quark mass slopes.
Our results are based on the determination of the $\mathrm{O}(a)$ improved PCAC masses via
\begin{align}
\label{eq:pcac-rat}
m(x_0) = \frac{\sdrv0 \fa^{ij}(x_0)+ac_\mathrm{A} \partial_0^\ast \partial_0^{}\fp^{ij}(x_0)}{2\fp^{ij}(x_0)}\,,
\end{align}
where $f_\mathrm{A}^{ij}$ and $f_\mathrm{P}^{ij}$ are Schrödinger functional correlation functions.
In order to improve the signal, these correlation functions are symmetrised with their $T$-symmetric counterparts  $\gA^{ij}(T-x_0)$ and $\gP^{ij}(T-x_0)$, which are constructed from the same operators $(\aimpr)^{ij}_0(x)$ and ${P}^{ij}(x)$ in the bulk but the pseudoscalar wall with operator $\mathcal{O^\prime}^{ji}$ positioned at the time boundary $x_0 = T$. For exact definitions see Appendix~\ref{app:sfcf}.
\begin{table}
	\centering
	\renewcommand{\arraystretch}{1.25}
	\setlength{\tabcolsep}{3pt}
	\begin{tabular}{lllll}
\toprule
 ID   & \multicolumn{2}{c}{$am$}   & $\zthree$                 & $\zfour$ \\ 
                       & $ \kappa^\mathrm{val}_{i}=\kappa_{i}^\mathrm{sea}$ &$\kappa^\mathrm{val}_i=\kappa_\mathrm{crit}$   \\
\midrule
 A3k3 & $\phantom{-}0.12143(82)$   & $\phantom{-}0.10759(77)$  &                       &                                                                                                    \\
 A3k1 & $\phantom{-}0.07316(184)$  & $\phantom{-}0.06440(193)$ &                       &                                                                                                    \\
 A3k2 & $\phantom{-}0.03070(85)$   & $\phantom{-}0.02588(89)$  &                       &                                                                                                    \\
 A3k6 & $\phantom{-}0.01246(54)$   & $\phantom{-}0.01029(54)$  &                       &                                                                                                    \\
 A3k4 & $\phantom{-}0.00465(56)$   & $\phantom{-}0.00370(57)$  &                       &                                                                                                    \\
 A1k3 & $\phantom{-}0.00095(93)$   & $\phantom{-}0.00074(93)$  & \phantom{-}0.8195(93) & \phantom{-}0.7454(94)                                                                              \\
 A1k4 & $-0.00119(33)$             & $-0.00100(33)$            & \phantom{-}0.8101(43) & \phantom{-}0.7520(58)                                                                              \\
 A1k1 & $-0.00287(61)$             & $-0.00229(61)$            & \phantom{-}0.7892(67) & \phantom{-}0.7189(79)                                                                              \\
 A3k5 & $-0.00952(50)$             & $-0.00864(49)$            &                       &                                                                                                    \\
      & $\phantom{-}0.0$           & $\phantom{-}0.0$          & \phantom{-}0.8184(77) & \phantom{-}0.7588(143)\\ \midrule %                                                                \\
 E2k1 & $\phantom{-}0.02083(19)$   & $\phantom{-}0.01117(27)$  &                       &                                                                                                    \\
 E2k2 & $\phantom{-}0.01072(16)$   & $\phantom{-}0.00592(17)$  &                       &                                                                                                    \\
 E1k1 & $\phantom{-}0.00265(22)$   & $\phantom{-}0.00153(23)$  & \phantom{-}0.8990(47) & \phantom{-}0.8619(54)                                                                              \\
 E1k2 & $-0.00022(19)$             & $-0.00017(19)$            & \phantom{-}0.8987(47) & \phantom{-}0.8580(64)                                                                              \\
      & $\phantom{-}0.0$           & $\phantom{-}0.0$          & \phantom{-}0.8987(43) & \phantom{-}0.8583(59)\\ \midrule %                                                                 \\
 B3k1 & $\phantom{-}0.01502(16)$   & $\phantom{-}0.00552(22)$  &                       &                                                                                                    \\
 B1k1 & $\phantom{-}0.00552(19)$   & $\phantom{-}0.00232(18)$  & \phantom{-}0.9972(45) & \phantom{-}0.9760(53)                                                                              \\
 B1k2 & $\phantom{-}0.00435(28)$   & $\phantom{-}0.00168(30)$  & \phantom{-}0.9963(73) & \phantom{-}0.9756(94)                                                                              \\
 B1k3 & $\phantom{-}0.00157(18)$   & $\phantom{-}0.00024(20)$  & \phantom{-}0.9839(48) & \phantom{-}0.9643(52)                                                                              \\
 B1k4 & $-0.00056(16)$             & $-0.00035(16)$            & \phantom{-}1.0004(50) & \phantom{-}0.9690(73)                                                                              \\
      & $\phantom{-}0.0$           & $\phantom{-}0.0$          & \phantom{-}0.9935(38) & \phantom{-}0.9654(50)\\ \midrule %                                                                 \\
 C1k1 & $\phantom{-}0.01322(17)$   & $\phantom{-}0.00304(21)$  & \phantom{-}1.0593(46) & \phantom{-}1.0446(42)                                                                              \\
 C1k2 & $\phantom{-}0.00601(11)$   & $\phantom{-}0.00148(11)$  & \phantom{-}1.0615(30) & \phantom{-}1.0517(35)                                                                              \\
 C1k3 & $-0.00110(11)$             & $-0.00029(11)$            & \phantom{-}1.0617(47) & \phantom{-}1.0542(42)                                                                              \\
      & $\phantom{-}0.0$           & $\phantom{-}0.0$          & \phantom{-}1.0621(36) & \phantom{-}1.0544(34)\\ \midrule %                                                                 \\
 D1k2 & $\phantom{-}0.00073(15)$   & $\phantom{-}0.00012(15)$  & \phantom{-}1.0896(89) & \phantom{-}1.0868(52)                                                                              \\
 D1k4 & $-0.00007(3)$              & $-0.00001(3)$             & \phantom{-}1.0908(12) & \phantom{-}1.0849(13)                                                                              \\
 D1k1 & $-0.00295(11)$             & $-0.00040(9)$             &                       &                                                                                                    \\
      & $\phantom{-}0.0$           & $\phantom{-}0.0$          & \phantom{-}1.0907(13) & \phantom{-}1.0850(12)                                                                              \\
\bottomrule
\end{tabular}

	\caption{For each ensemble, identified in the first column by an ID label, we list our results for the PCAC mass $am$ for simulations with
		$\kappa^\mathrm{val} = \kappa^\mathrm{sea}$ (second column) and $\kappa^\mathrm{sea} \neq \kappa^\mathrm{val} = \kappa_\mathrm{crit}$ (third column). The last two columns contain $Z$ results obtained from the Ward identity~(\ref{eq:WIf}). The final results are those extrapolated to the chiral limit at each  $\beta=6/g_0^2$ (last line of each data grouping). The labels $\zthree$ and $\zfour$ refer to different choices of time slices with operator insertions in the correlation functions (see text for details).}
	\label{table_am_Z}
\end{table}

We first determine the required correlation functions in a unitary setup, $\kappa^\mathrm{val} = \kappa^\mathrm{sea}$. From these we can obtain $\kappa_\mathrm{crit}$ as will be detailed below. In a second step we compute the same correlation functions in a non-unitary setup where $\kappa^\mathrm{sea} \neq \kappa^\mathrm{val} = \kappa_\mathrm{crit}$.
In Fig.~\ref{pcac_plateau} we show the temporal dependence of the current quark mass $\m(x_0)$ for both of these setups for the representative ensembles B3k1 and D1k2 and demonstrate that they form well-defined plateaux as a function of time, away from the Dirichlet boundaries.
Our final estimate for the PCAC masses is obtained by averaging $m(x_0)$ over the central third of the temporal extent of the lattice.
This choice is motivated by the coarsest lattices; the plateaux for the finer ones also extend closer to the boundary before lattice artefacts become relevant as can be seen in Fig.~\ref{pcac_plateau}.
The plateau range is adapted according to the time extent for each value of $\beta$, so as to preserve the line of constant physics.
Our PCAC mass estimates in both setups are listed in Table~\ref{table_am_Z} for all ensembles.

\begin{figure}
	\centering
	\vspace{-0.15cm}
	\begin{subfigure}[b]{1.0\textwidth}
		\centering
		\includegraphics[width=\textwidth]{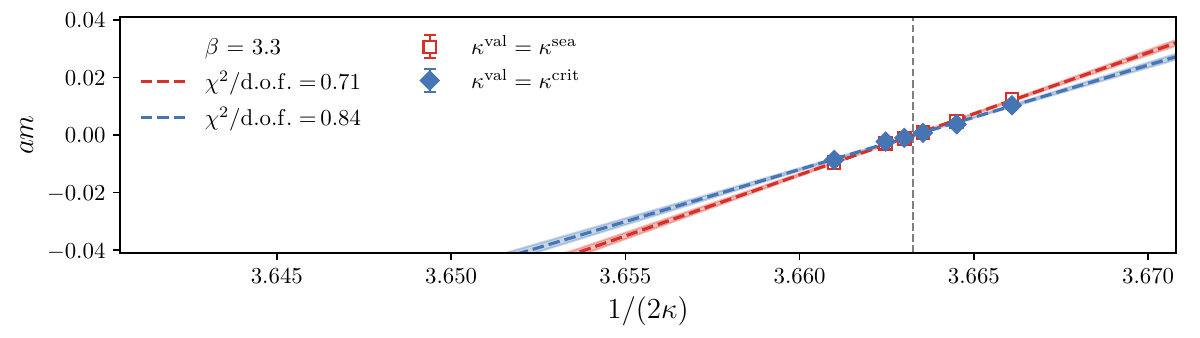}
	\end{subfigure}
	\vspace{-0.15cm}
	\begin{subfigure}[b]{1.0\textwidth}
		\centering
		\includegraphics[width=\textwidth]{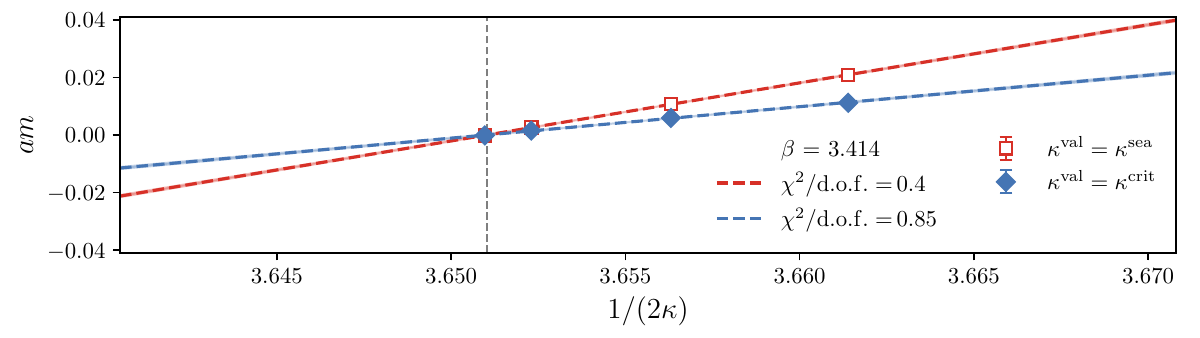}
	\end{subfigure}
	\vspace{-0.15cm}
	\begin{subfigure}[b]{1.0\textwidth}
		\centering
		\includegraphics[width=\textwidth]{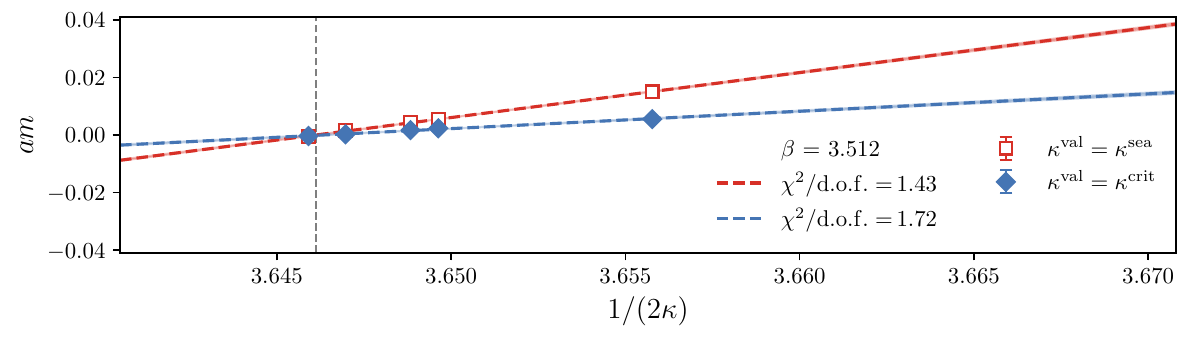}
	\end{subfigure}
	\vspace{-0.15cm}
	\begin{subfigure}[b]{1.0\textwidth}
		\centering
		\includegraphics[width=\textwidth]{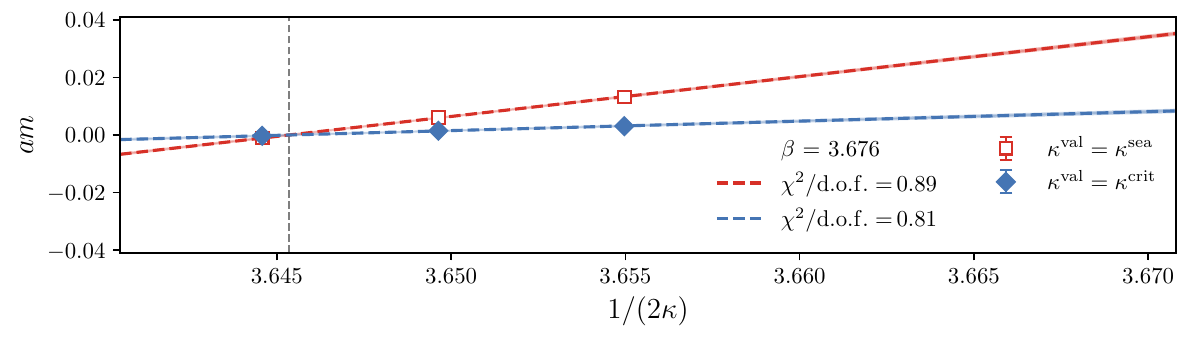}
	\end{subfigure}
	\vspace{-0.15cm}
	\begin{subfigure}[b]{1.0\textwidth}
		\centering
		\includegraphics[width=\textwidth]{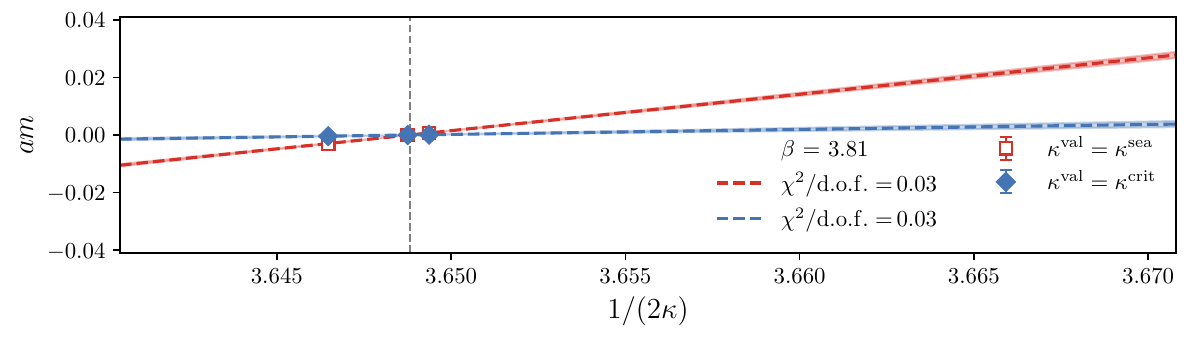}
	\end{subfigure}
	\caption{PCAC masses $am$ fitted linearly in $1/(2\kappa)$, for all simulated $\beta$ values (i.e. for decreasing lattice spacings from top to bottom). Open squares and filled diamonds are results in the unitary and non-unitary setups, respectively. Note that horizontal and vertical axes are identical for all values of $\beta$, so as to highlight the different ranges of $\kappa$ and the change of $\kappa_\mathrm{crit}$ marked by the vertical dashed lines.}
	\label{fig:mass-vs-kappainv}
\end{figure}

In order to extract $Zr_\mathrm{m}$ and $Z(r_\mathrm{m}-1)$ from the slopes of the current quark masses with respect to the bare quark masses, we plot $am$ against the inverse hopping parameter $1/(2\kappa)$ for both the unitary and the non-unitary setup, as demonstrated in Fig.~\ref{fig:mass-vs-kappainv}. 
We generally observe that $m$ behaves linearly as a function of $1/(2\kappa)$ in the range $-0.1 \lesssim L m \lesssim 0.3$.
For the ensembles A3k1, A3k2, and A3k3 (not displayed in Fig.~\ref{fig:mass-vs-kappainv}), which correspond to $Lm \gtrsim 0.3$, linearity is lost.
\begin{figure}
	\centering
	\includegraphics[width=1.0\textwidth]{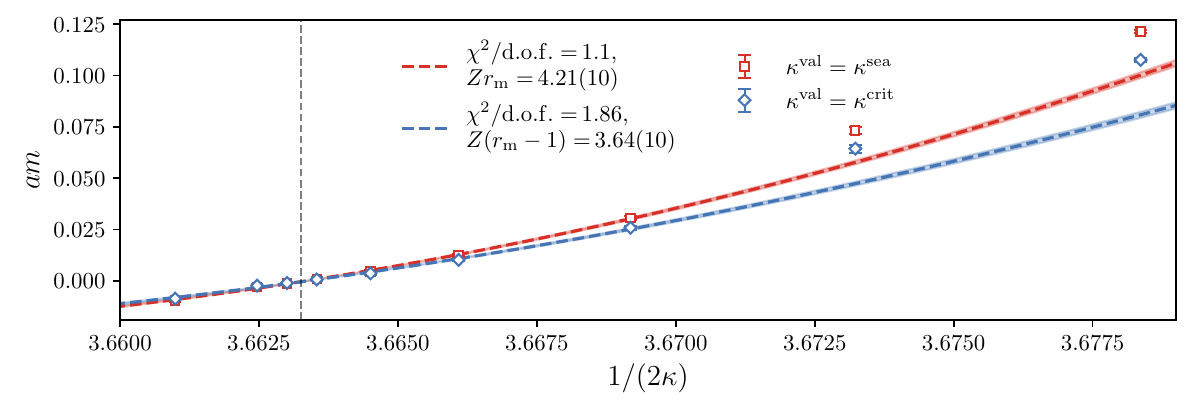}
	\caption{PCAC masses $am$, at $\beta=3.3$, fitted with a quadratic polynomial in $1/(2\kappa)$. Squares and diamonds are results in the unitary and non-unitary setups, respectively. Note that the two rightmost points (A3k1, A3k3) are not included in the fit, while A3k2 (at $1/(2\kappa) = 3.6692$) is. The vertical dashed line is positioned at $\kappa_\mathrm{crit}$ from the linear fit; see Table~\ref{table_slopes}. The $Zr_\mathrm{m}$ and $Z(r_\mathrm{m}-1)$ values shown in the legend are obtained from the linear term of the quadratic polynomial.}
	\label{fig:non-linear-fit}
\end{figure}
Results from these ensembles have thus not been included in the linear fits. The good linear behaviour of the data from the remaining ensembles is justified {\it a posteriori}, by the small $\chi^2/\mathrm{d.o.f}$. of our fits, as shown in Fig.~\ref{fig:mass-vs-kappainv}.

We also probe the non-linear regime in both setups for $\beta=3.3$ by performing a quadratic fit, in the presence of the ensembles A3k1, A3k2, and A3k3, as displayed in Fig.~\ref{fig:non-linear-fit}. For both setups, fits confirm the presence of $\rmO((am_\mathrm{q})^2)$ effects in this case. The two rightmost points (A3k1, A3k3) have not been included in these fits. Including them would result in a very large value of $\chi^2/\mathrm{d.o.f}$. This may also be related to the fact that no clear-cut plateaux are seen in the current quark mass data for these ensembles. This could be explained by the fact that (boundary) cut-off effects for these comparatively large masses (in lattice units) are substantial. 
Estimates of $Zr_\mathrm{m}$ and $Z(r_\mathrm{m}-1)$, obtained as the linear coefficient of the quadratic fits around $\kappa_\mathrm{crit}$, are compatible with those from linear fits. The influence of the quadratic term on our final result is therefore negligible. This ensures that our results are not affected by $\rmO((a \mq)^2)$ systematic errors at $\beta=3.3$, which is our coarsest lattice. The same conclusion holds for the finer lattices, since also for them $a \mq$ is small and linear fits have small $\chi^2/\mathrm{d.o.f.}$

As implied by eq.~(\ref{eq:mPCAC-msub-2}), $\kappa_\mathrm{crit}$ and $Zr_\mathrm{m}$ are assessed as the intercept and the slope of the linear fit to the unitary data. Similarly, eq.~(\ref{eq:mPCAC-msub-3}) tells us that $Z(r_\mathrm{m}-1)$ can be estimated from the slope of the linear fit to the non-unitary data. Our final findings for $Zr_\mathrm{m}$, $Z(r_\mathrm{m}-1)$, and $\kappa_\mathrm{crit}$ are listed in Table~\ref{table_slopes}.
\begin{table}[h!]
	\centering
	\renewcommand{\arraystretch}{1.25}
	\setlength{\tabcolsep}{3pt}
	\begin{tabular}{llll}
\toprule
 $\beta$   & $Zr_\mathrm{m}$   & $Z(r_\mathrm{m}-1)$   & $\kappa_\mathrm{crit}$   \\
\midrule
 3.3       & 4.240(134)        & 3.621(133)            & 0.1364904(18)            \\
 3.414     & 2.015(24)         & 1.092(29)             & 0.1369478(26)            \\
 3.512     & 1.561(21)         & 0.603(26)             & 0.1371320(26)            \\
 3.676     & 1.383(19)         & 0.329(20)             & 0.1371611(25)            \\
 3.81      & 1.263(47)         & 0.173(40)             & 0.1370310(9)             \\
\bottomrule
\end{tabular}

	\caption{Results from the PCAC mass analyses. The second and fourth column show results obtained in a unitary setup; the third column refers to the non-unitary setup.}
	\label{table_slopes}
\end{table}

\subsection{\texorpdfstring{Renormalisation constant $Z$}{Renormalisation constant Z}}
\label{sec:z}
As the next step in our analysis, we extract the renormalisation constant $Z \equiv (\zp/\zs\za)$ from the ratio~(\ref{eq:WIf}), using the subset of gauge field ensembles listed in Table~\ref{tab:sim_table} which are {\it not} emphasised in italic font.\footnote{As explained in Section~\ref{sec:setup}, the ensembles in \textit{italics} have been generated for the purpose of performing reliable fits of the data in Fig.~\ref{fig:mass-vs-kappainv} and \ref{fig:non-linear-fit}, in order to accurately measure their slopes. These extra ensembles have not been used for the computation of $Z $, as they do not increase the accuracy of the result. D1k1 (marked by an asterisk) is also not taken into account.} The correlation functions in eq.~(\ref{eq:WIf}) are computed for two choices of $t_1$ and $t_2$. Our first choice is $t_1 \approx T/3$ and $t_2 \approx 2T/3$, and the results obtained in this fashion are denoted as $\zthree$. Alternatively, choosing $t_1 \approx T/4$ and $t_2 \approx 3T/4$ yields a second $Z$ estimate denoted as $\zfour$. When $T/3$ and $T/4$ are not integers, $t_1$ and $t_2$ are rounded up/down to the nearest integer.

\begin{figure}
	\centering
	\begin{subfigure}[b]{0.49\textwidth}
		\centering
		\includegraphics[width=1.0\textwidth]{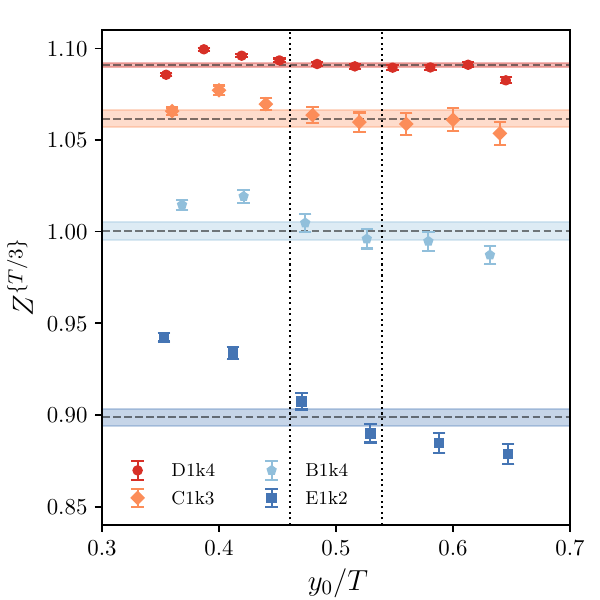}
	\end{subfigure}
	\begin{subfigure}[b]{0.49\textwidth}
		\centering
		\includegraphics[width=1.0\textwidth]{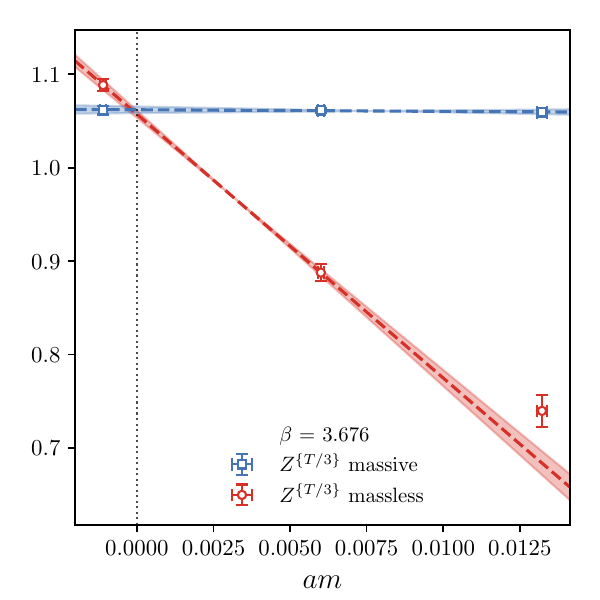}
	\end{subfigure}
	\caption{ \textit{Left:} Ward identity estimates of $Z$, plotted against time $y_0/T$, for one representative ensemble for each lattice spacing (except for $						\beta=3.3$, corresponding to the coarsest lattice). The dashed vertical lines bracket the two central time slices that determine the final value of $Z$.
		\textit{Right:} Chiral extrapolation of $Z$ at fixed $\beta$ obtained from the Ward identity with  the mass term (squares) and without it (diamonds). In the massless case, a possible linear range in $am$ is illustrated by the dashed line joining the two leftmost points.
		In the massive case, no significant quark mass dependence is observed; the dashed line through the squares is a linear fit where the slope vanishes within its uncertainty. Note that the errors of the PCAC masses are also displayed and taken into account in the fits via orthogonal distance regression \cite{Boggs1989}.}
	\label{z_plateau}
\end{figure}

In the left part of Fig.~\ref{z_plateau}, we depict $\zthree$ as a function of $y_0/T$ for several representative ensembles (we remind the reader that $T$ is approximately constant in physical units).
Contrary to the PCAC masses in Fig.~\ref{pcac_plateau}, these local estimators of $Z$ do not exhibit plateau-like behaviour; this was also observed for a similar Ward identity adopted to compute the improvement coefficient of the vector current in Ref.~\cite{Heitger:2020zaq}. Note, however, that this is not problematic; since $Z$ is obtained from a Ward identity, its value at any time slice qualifies as a well-defined estimate.
We prefer to err on the side of caution and quote the average of the two central time slices as our best $Z$ estimate. Results for the two determinations of $Z$ are collected in Table~\ref{table_am_Z}, where we see that $\zthree$ and $\zfour$ are not compatible, indicating the presence of lattice artefacts that also differ noticeably. We consider $\zthree$ the more reliable estimate because the operator insertions in this case, being further from $x_0 = 0$ and $x_0 = T$, are expected to lead to less contamination through cut-off effects induced by the boundaries.

Since the Ward identity~(\ref{eq:WIf}) is only valid up to lattice artefacts of $\mathrm{O}(am,a^2)$, we have to interpolate our data to the chiral point, in order to eliminate the $\mathrm{O}(am)$-effects and be left with $\mathrm{O}(a^2)$ only. As an additional cross-check we also compute $Z$ without the ``mass term'' $2 am \ftps(t_2,t_1)$ in the Ward identity~(\ref{eq:WIf}), where $am$ is the PCAC mass from the unitary setup discussed in the previous section.
This chiral interpolation is demonstrated for $\beta=3.676$ in the right part of Fig.~\ref{z_plateau}. While the data including the ``mass term'' shows a very flat behaviour with respect to the current quark mass (where the associated fit parameter even vanishes within its uncertainty except for the coarsest lattice spacing), the truncated Ward identity results in a considerably larger slope. If we exclude the rightmost data point for the identity without the ``mass term'', linear fits to both datasets still agree in the chiral limit.
This situation resembles closely what was observed in Ref.~\cite{Heitger:2020mkp}, where $Z$ was measured employing a different Ward identity. 
We note that the linear fit is based on the orthogonal distance regression method \cite{Boggs1989}, taking into account both the error of dependent and independent variables.
The final results for $\zthree$ and $\zfour$ at the chiral point are also listed in Table~\ref{table_am_Z}. Compared to the indirect Ward identity determination of Ref.~\cite{Heitger:2020mkp}, they have considerably smaller errors. This confirms the expectation that the simpler structure of the correlation functions building the Ward identity (\ref{eq:SPImassintlatt}) is preferable from a numerical perspective; see the discussion at the end of Section~\ref{sec:ward_identity}.
On the other hand, compared to the so-called 'LCP-0' determination of Ref.~\cite{deDivitiis:2019xla}, our results are of similar accuracy across the bare couplings investigated. We will use our results (Table~\ref{table_am_Z}) for a precise estimation of $r_\mathrm{m}$ in the following.
More details on the relative cut-off effects between the present determination of $Z$ and the results obtained in Refs.~\cite{deDivitiis:2019xla,Heitger:2020mkp,Bali:2016umi} can be found in Appendix~\ref{app:z}.

\subsection{\texorpdfstring{Results for $r_\mathrm{m}$}{Results for rm}}
\label{sec:rm-results}	
In the final step of our analysis we combine the values of $Zr_\mathrm{m}$ obtained in a unitary setup, $Z(r_\mathrm{m}-1)$ in a non-unitary setup, and $Z$ from a chiral Ward identity, in order to arrive at different estimates for $r_\mathrm{m}$.
Combining the first two, we construct $\rmonetwo$, defined as
\begin{align}
\rmonetwo &= \bigg(1-\left[\frac{Z(r_\text{m}-1)}{Zr_\text{m}}\right]\bigg)^{-1}
\label{rmonetwo}\,,
\intertext{where the superscripts ``u'' and ``nu'' stand for ``unitary'' and ``non-unitary'', respectively. Combining $Zr_\mathrm{m}$ and $Z$, results in $\rmonethree$, defined as}
\rmonethree &= \frac{Zr_\text{m}}{Z}
\label{rmonethree}\,.
\intertext{As mentioned above, this comes in two versions, $\rmonethreethree$ and  $\rmonethreefour$. Moreover, from the second and third result we gain $\rmtwothree$ given by}
\rmtwothree &= \frac{Z(r_\text{m}-1)}{Z} + 1
\label{rmtwothree}\,,
\end{align}
which is again worked out for two cases,  $\rmtwothreethree$ and $\rmtwothreefour$.
All our results for $r_\mathrm{m}$ from these different determinations just outlined are gathered in Table~\ref{table_rm_12_T3}. 

\begin{table}
	\centering
	\renewcommand{\arraystretch}{1.4}
	\setlength{\tabcolsep}{3pt}
	\begin{tabular}{llllll}
\toprule
 $\beta$   & $\rmonetwo$   & $\rmonethreethree$   & $\rmtwothreethree$   & $\rmonethreefour$   & $\rmtwothreefour$   \\
\midrule
 3.3       & 6.848(569)    & 5.181(172)           & 5.424(169)           & 5.588(207)          & 5.772(199)          \\
 3.414     & 2.183(44)     & 2.242(26)            & 2.215(32)            & 2.348(32)           & 2.272(35)           \\
 3.512     & 1.629(32)     & 1.571(21)            & 1.607(26)            & 1.617(22)           & 1.625(27)           \\
 3.676     & 1.312(20)     & 1.303(15)            & 1.309(19)            & 1.312(16)           & 1.312(19)           \\
 3.81      & 1.158(37)     & 1.158(44)            & 1.158(37)            & 1.164(44)           & 1.159(37)           \\
\bottomrule
\end{tabular}

	\caption{Results for $r_\mathrm{m}$, obtained via eqs.~(\ref{rmonetwo}) to (\ref{rmtwothree}).}
	\label{table_rm_12_T3}
\end{table}

\begin{figure}[t]
	\centering
	\begin{subfigure}[b]{0.49\textwidth}
	\centering
	\includegraphics[width=1.0\textwidth]{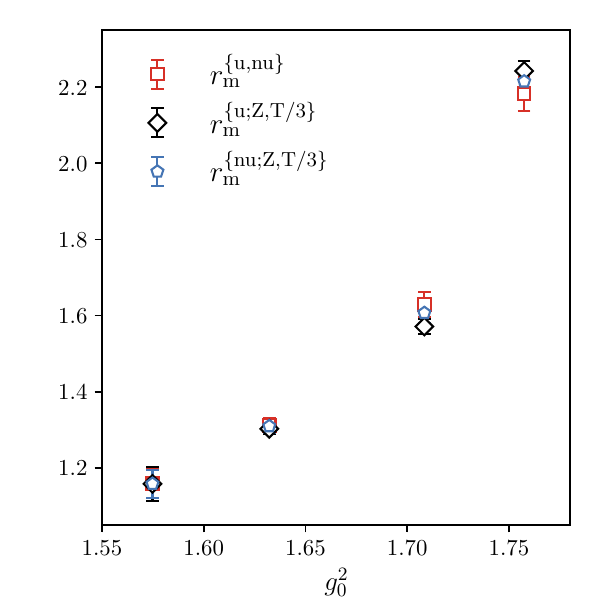}
	\end{subfigure}
	\begin{subfigure}[b]{0.49\textwidth}
		\centering
		\includegraphics[width=1.0\textwidth]{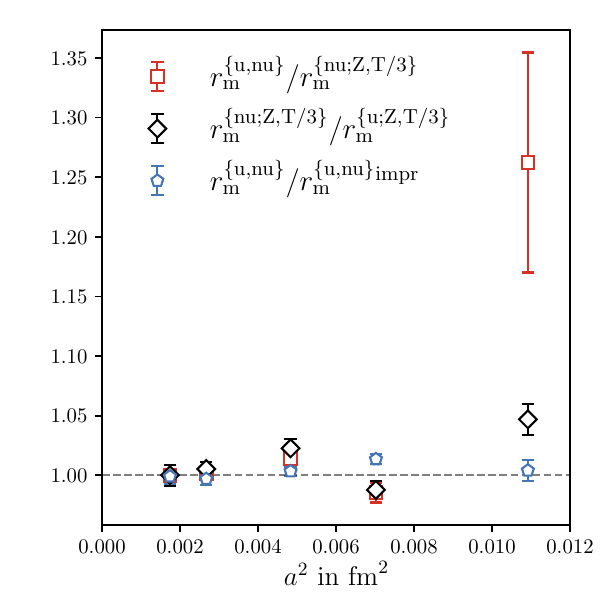}
	\end{subfigure}
	\caption{\textit{Left:} Results for different $r_\mathrm{m}$ estimates as reported in Table~\ref{table_rm_12_T3}. The results for $\beta=3.3$ are not shown. \textit{Right:} Ratio of different $r_\mathrm{m}$ determinations as a function of the squared lattice spacing. The dashed horizontal line indicates the expected continuum result.}
	\label{rm_plot}
\end{figure}

In principle, the different estimates can differ by $\mathrm{O}(a^2)$ ambiguities. In Fig.~\ref{rm_plot} (left) the three determinations $\rmonetwo$, $\rmonethreethree$, and $\rmtwothreethree$ are plotted against the bare coupling squared; to be able to distinguish between the different estimates, the data points corresponding to the coarsest lattice spacing ($\beta=3.3$) are omitted as they exhibit large cut-off effects and are thus well out of the range displayed here. Results are compatible within their respective $1\sigma$-errors. In Fig.~\ref{rm_plot} (right) we take a closer look at this behaviour by plotting ratios of different $r_\mathrm{m}$ estimates as functions of the lattice spacing squared; the corresponding lattice spacings can be found in Table~\ref{tab:sim_table}. Since the ratios have been computed on a line of constant physics, and assuming that we are in a scaling region where Symazik's effective theory of cut-off effects applies, they are expected to be polynomials in the lattice spacing, tending to 1 in the continuum limit. In this context we introduce an additional determination, $r_\mathrm{m}^{\{ \mathrm{u},\mathrm{nu};\mathrm{impr}\}}$, which only differs from $\rmonetwo$ by an improved version of the derivative 
$\sdrv0$ in eq.~(\ref{eq:pcac-rat}).\footnote{The improved derivative is defined as $a\partial_\mu f(x) \equiv \frac{1}{12}[-f(x+2a\hat\mu)+8f(x+a\hat\mu) - 8f(x-a\hat\mu)+f(x-2a\hat\mu)]$ and its corresponding second derivative by $a^2\partial_\mu^*\partial_\mu f(x) \equiv \frac{1}{12}[-f(x+2a\hat\mu)+16f(x+a\hat\mu)-30f(x) +16f(x-a\hat\mu)-f(x-2a\hat\mu)]$ as shown by eq.~(B.4) in Ref.~\cite{deDivitiis:2019xla}. } These ratios are very close to one except for one of the data points at $\beta=3.3$, for which the ratio is significantly larger.
 Even though it would be sufficient to demonstrate that these ratios of $r_\mathrm{m}$ approach unity with a rate $\propto a^2$ or higher in our particular line of constant physics framework,
  such ambiguities appear to be nearly absent for $a<0.1 \mathrm{fm}$.
 
We tried to model the data sets with and without the $\beta=3.3$ points, using polynomials in the lattice spacing, constrained to one in the continuum limit. When a linear term is included, we obtain unsatisfactory fits with $\chi^2/\mathrm{d.o.f.}>3$. We thus conclude that our results are compatible with the theoretical expectation of $\mathrm{O}(a^2)$ lattice artefacts or higher (see also Appendix~\ref{app:z}).

\begin{figure}[t]
	\centering
	\includegraphics[width=1.0\textwidth]{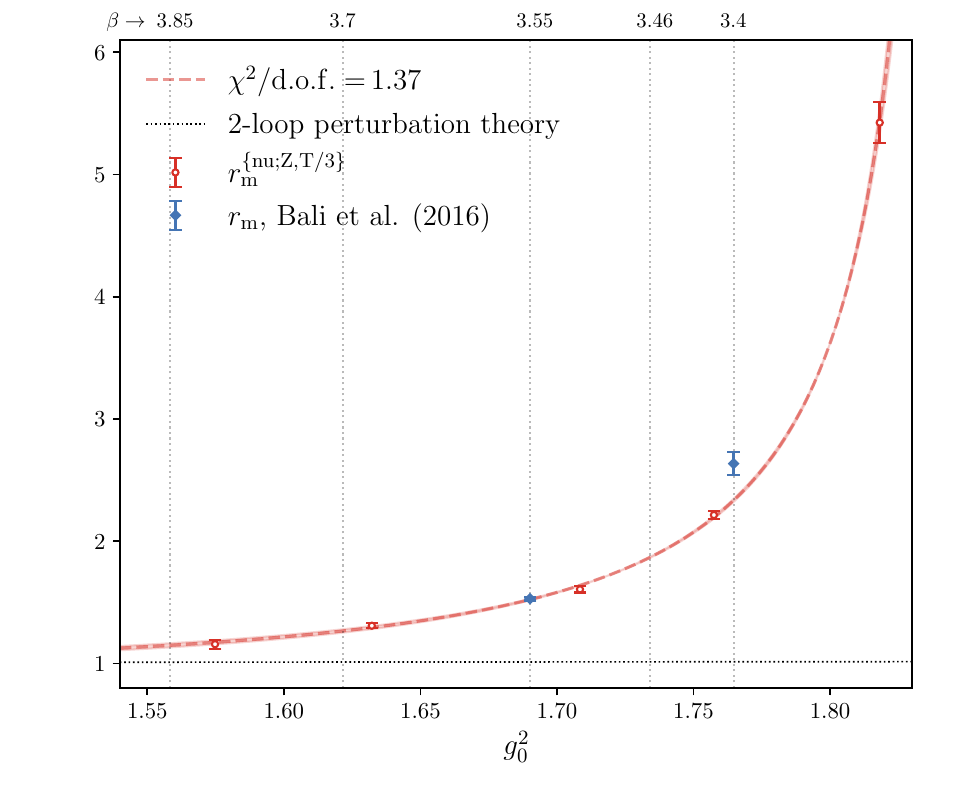}
	\caption{Non-perturbative determination of $\rmtwothreethree$ (open circles), compared to the results of Ref.~\cite{Bali:2016umi} (filled diamonds) and those of two-loop perturbation theory~\cite{Constantinou:2016ieh} (horizontal dotted line). The dashed line is the interpolation~(\ref{eq:fit}) and the vertical dotted lines correspond to the bare couplings used in \textit{CLS} simulations.}
	\label{rm_interpolation}
\end{figure}

As our preferred determination of $r_\mathrm{m}$ we advocate $\rmtwothreethree$ because of its small statistical errors in our range of bare couplings and the poorer scaling behaviour of the other estimators at the coarsest lattice spacing.
In Fig.~\ref{rm_interpolation} we show this result including the two-loop perturbative prediction of Ref.~\cite{Constantinou:2016ieh}. An important observation is that the non-perturbative estimates strongly deviate from the perturbative prediction in this region of strong couplings. A similar behaviour was also observed in several studies of renormalisation factors for which one-loop perturbative predictions are available (see, e.g. \cite{DallaBrida:2018tpn,Heitger:2020zaq}). Here, we confirm this finding also for two-loop perturbation theory. We also compare our results with those of other works. In Ref.~\cite{Bali:2016umi}, $r_\mathrm{m}$ was determined for two values of the bare coupling, from an alternative renormalization condition. As inferred by Fig.~\ref{rm_interpolation} this result agrees with ours at the smaller coupling, while it deviates notably at the larger coupling, most likely due to $\mathrm{O}(a^2)$ ambiguities (or higher).

Our final result consists of a continuous interpolation formula for $r_\mathrm{m} = \rmtwothreethree$.
Our data is best described by a Pad\'e ansatz, constrained to the two-loop prediction of Ref.~\cite{Constantinou:2016ieh} for small couplings, of the form
\begin{subequations}
	\label{eq:fit}
	\begin{equation}
	r_\text{m}(g_0^2)=1.0 +  0.004630 \, g_0^4\times\left\{\frac{ 1 + c_1\, g_0^2 + c_2\, g_0^4}{1+c_3\,g_0^2}\right\}\,,
	\end{equation}
where
	\begin{equation}
	\quad c_i = \left(-7.86078, 5.49175, -0.54078\right)\,,
	\end{equation}
and
	\begin{equation}
	\mathrm{cov}(c_i,c_j) = \left(\begin{array}{lll}
	\phantom{-}3.699770\phantom{ \times 10^{-1}}&-2.197442\phantom{ \times 10^{-1}}&-1.147397 \times 10^{-3}\\-2.197442\phantom{ \times 10^{-1}}&\phantom{-}1.306516\phantom{ \times 10^{-1}}&\phantom{-}6.951628 \times 10^{-4}\\-1.147397 \times 10^{-3}&\phantom{-}6.951628 \times 10^{-4}&\phantom{-}5.895744 \times 10^{-7} % corrected on arxiv in v3
	\end{array}\right)\,,
	\end{equation}
\end{subequations}
which is also displayed in Fig.~\ref{rm_interpolation}. The fit function describes our data with $\chi^2/{\rm d.o.f.}=1.37$ and provides errors of a size comparable to the fitted data points. 

The interpolation formula can now be used in order to determine $r_\mathrm{m}$ at the couplings used in \textit{CLS} simulations for the computation of hadronic quantities~\cite{Bruno:2014jqa,Bali:2016umi,Mohler:2017wnb}. 
Since the \textit{CLS} coupling $\beta=3.85$ lies outside the range of our $r_\mathrm{m}$ computations, we perform a short extrapolation in order to provide a value for $r_\mathrm{m}(\beta=3.85)$. 
A systematic error, estimated as the difference between the lower error bar of our data point at $\beta=3.81$ and the extrapolated value at $\beta=3.85$ ($\sigma_\mathrm{syst}= 0.027$) is added to the statistical error ($\sigma_\mathrm{stat}=0.018$) in quadrature. Our final  $r_\mathrm{m}$ results at the \textit{CLS} couplings are collected in Table~\ref{tab:rm_cls}.

\begin{table}[h]
	\centering
	\renewcommand{\arraystretch}{1.4}
	\setlength{\tabcolsep}{3pt}
	\begin{tabular}{llllll}
\toprule
 $\beta$                 & 3.4       & 3.46      & 3.55      & 3.7       & 3.85              \\
 \midrule $r_\mathrm{m}$ & 2.335(31) & 1.869(19) & 1.523(14) & 1.267(16) & 1.149(18)(27)[33] \\
\bottomrule
\end{tabular}

	\caption{Values for $r_\mathrm{m}$ at the couplings used in \textit{CLS} simulations, obtained from the interpolation formula~(\ref{eq:fit}). As mentioned in the text, an additional systematic error was added to the $\beta=3.85$ result. The errors are displayed in this way:  $(\sigma_\mathrm{stat})(\sigma_\mathrm{syst})[\sigma_\mathrm{total}]$.}
	\label{tab:rm_cls}
\end{table}

\section{Summary}
\label{sec:conclusions}
With the non-perturbative computation of the ratio of the renormalisation
constants of non-singlet and singlet scalar densities,
$\rmsea \equiv \zs/\zss$, presented in this paper we have addressed a
quantity, which not only enters the renormalisation pattern of quark masses
in lattice QCD with Wilson fermions, but also constitutes an important
ingredient in calculations of renormalised nucleon (and other baryon) matrix
elements of singlet scalar densities, known as sigma terms.

Our strategy to calculate $\rmsea$ merges the functional dependences of
the PCAC quark mass in terms of the subtracted quark mass, evaluated in a
unitary as well as a non-unitary setting with respect to the choice of sea and
valence quark masses.
In the vicinity of the chiral limit, these dependences are found to be
linear, so that $\rmsea$ can be obtained through the associated quark mass
slopes with confidence and superior control of statistical and
systematic errors.
The finite-volume numerical simulations of $\rmO(a)$ improved QCD with
Schr\"odinger functional boundary conditions that enter the analysis
realise a line of constant physics by working in a volume of spatial extent
$L\approx 1.2\,$fm and thereby fixing all other relevant length scales
in physical units.
This guarantees that $\rmsea$ becomes a smooth function of the bare gauge
coupling as the lattice spacing is varied, where any potentially remaining
intrinsic ambiguities disappear monotonically towards the continuum limit at
a rate that stays beyond the sensitivity of the $\rmO(a)$ improved theory.

Our central results, which hold for a lattice discretisation of QCD with
three flavours of non-perturbatively $\rmO(a)$ improved Wilson-clover sea
quarks and tree-level Symanzik-improved gluons, are the continuous
parameterisation of $\rmsea$ as a function of the squared bare gauge
coupling $g_0^2=6/\beta$ in eq.~(\ref{eq:fit}), as well as its values in
Table~\ref{tab:rm_cls} at the specific strong-coupling $\beta$ values of
large-volume \textit{CLS}
simulations~\cite{Bruno:2014jqa,Bruno:2016plf,Bali:2016umi,Mohler:2017wnb}.

Along with the numerical implementation of our strategy to extract
$\rmsea$, we have also developed a new method to determine the scale
independent combination $Z=\zp/(\zs\za)$ of renormalisation parameters of
quark bilinears in the pseudoscalar, (non-singlet) scalar and axial vector
channel, respectively.
It relies upon a Ward identity that, according to our knowledge,
has not yet appeared explicitly in the literature.
Since, as explained in Sections~\ref{sec:masses-rm-Z} and
\ref{sec:ward_identity}, the renormalisation factor $Z$ is actually
required to isolate $\rmsea$ from the unitary and
non-unitary quark mass slopes, we have employed the estimates on $Z$ from
this approach in our final results of $\rmsea$.
However, this was primarily done for practical reasons and served the purpose
of demonstrating the feasibility of the Ward identity method for $Z$.
In fact, it is apparent from the discussion in Appendix~\ref{app:z}
and Figure~\ref{plot_z} that these new values for $Z$ are fully compatible
with the earlier determinations available from
Refs.~\cite{deDivitiis:2019xla,Heitger:2020mkp} and are neither superior in statistical precision nor in systematics regarding lattice artefacts. Nevertheless we give an interpolation formula for the present $Z$ (Table ~\ref{tab:Z_comparison}) for completeness.

Finally we recall the subtlety discussed in Section~\ref{sec:masses-rm-Z}: away from the chiral limit, the dependence of (re)normalisation parameters should be $Z(\tilde g_0^2), \rmsea(\tilde g_0^2)$, with $\tilde g_0^2$ defined in eq.~(\ref{eq:g0-tilde}). In order to be able to combine our results with CLS low-energy quantities such as those of Refs.~\cite{Bruno:2016plf,Bruno:2019vup}, we should use the expansion
\begin{equation}
\label{eq:Zg0-tilde}
Z(\tilde g_0^2) = Z(\gosq) \Big [ 1 + \dfrac{\partial \ln Z(g_0^2)}{\partial \gosq} \dfrac{1}{\Nf}b_g(\gosq) \gosq a  \Tr \Mq \Big ] \, ,
\end{equation} 
see also  Ref.~\cite{Gerardin:2018kpy}, and similarly for $\rmsea(\tilde g_0^2)$. At present, $b_g$ is only known in perturbation theory~\cite{Sint:1997jx}; $b_g = 0.012 \Nf g_0^2$. The correction $\partial \ln Z/\partial \gosq$ as well as $\partial \ln \rmsea/\partial \gosq$ , computed at CLS values of the inverse coupling $\beta$, can be found in Table~\ref{tab:z_cls}.

\begin{acknowledgement}%
This work is supported by the Deutsche Forschungsgemeinschaft (DFG) through the Research Training Group ``GRK 2149: Strong and Weak Interactions -- from Hadrons to Dark Matter'' (J. H., F. J. and P. L. J. P.). We acknowledge the computer resources provided by the WWU IT, formerly ‘Zentrum für Informationsverarbeitung (ZIV)’, of the University of Münster (PALMA-II HPC cluster) and thank its staff for support.

\end{acknowledgement}

\clearpage

\begin{appendix}
\section{Schr\"odinger functional correlation functions}\label{app:sfcf}
\newcommand{\braketf}[1]{\left\langle#1\right\rangle}
The Schr\"odinger functional correlation functions employed in this work are defined as
\begin{align}
f_\mathrm{P}^{ij} & = -\frac{1}{2} \frac{a^9}{L^3} \sum_{\bf x, u, v} \braketf{\bar\psi_i(x) \gamma_5 \psi_j(x) \cdot \bar\zeta_j({\bf v}) \gamma_5 \zeta_i(\bf u)}\,, \label{fp} \\
g_\mathrm{P}^{ij} & = -\frac{1}{2} \frac{a^9}{L^3} \sum_{\bf x, u, v} \braketf{\bar\psi_i(x) \gamma_5 \psi_j(x) \cdot \bar\zeta'_j({\bf u}) \gamma_5 \zeta'_i(\bf v)}\,, \\
f_\mathrm{A}^{ij} & = -\frac{1}{2} \frac{a^9}{L^3} \sum_{\bf x, u, v} \braketf{\bar\psi_i(x) \gamma_0 \gamma_5 \psi_j(x) \cdot \bar\zeta_j({\bf u}) \gamma_5 \zeta_i(\bf v)}\,, \\
g_\mathrm{A}^{ij} & = -\frac{1}{2} \frac{a^9}{L^3} \sum_{\bf x, u, v} \braketf{\bar\psi_i(x) \gamma_0 \gamma_5 \psi_j(x) \cdot \bar\zeta'_j({\bf u}) \gamma_5 \zeta'_i(\bf v)}\,,\\
F_1^{ij} & = -\frac{1}{2} \frac{a^{12}}{L^6} \sum_{\bf u', v', u, v} \braketf{\bar\zeta'_i({\bf u'}) \gamma_5 \zeta'_j({\bf v'}) \cdot \bar\zeta_j({\bf u}) \gamma_5 \zeta_i(\bf v)}\,.
\end{align}
They refer to the general case of two distinct, i.e. not necessarily mass-degenerate quark flavours $i,j$. Summation over the indices $i$ and $j$ is not implied.
The space-time point $x$ lies in the lattice bulk; i.e. $0 < x_0 < T$. The Dirichlet boundary fields $\bar\zeta_j(\bf u)$ and $\zeta_i(\bf v)$ live on time slice $x_0=0$, while $\bar\zeta'_j(\bf u')$ and $\zeta'_i(\bf v')$ live on time slice $x_0=T$; the boundary fields are introduced in Ref.\cite{Luscher:1996sc}.
\section{Wick contractions of correlation functions}
\label{app:corr-funct}

In this appendix we briefly explain how to obtain eq.~(\ref{eq:WIf}) from eq.~(\ref{eq:SPImassintlatt}).
The idea is to perform the Wick contractions of the correlation functions, arriving at expressions
which are traces of flavour matrices, multiplying traces of products of quark propagators and $\gamma$-matrices. This procedure has been described in full detail in Ref.~\cite{Heitger:2020mkp}, which deals with more complicated Ward identities; we refer the reader to that work for unexplained notation. Here we will only present the main features of the proof.

We start with the r.h.s. of eq.~(\ref{eq:SPImassintlatt}). The Wick contractions result in
\begin{align}
- d^{abe} \, a^3 \sum_{\bf y} \langle  P^e(y) \cO^c \rangle
=& - d^{abe} \Tr[T^e T^c] \dfrac{a^{9}}{L^3} \sum_{\bf y} \sum_{\bf u,v} \Bigg \langle \tr \bigg \{ [\psi(y) \bar \zeta({\bf u})]_{\rm F} \gamma_5 [\zeta({\bf v}) \bar \psi(y)]_{\rm F} \gamma_5 \bigg \} \Bigg \rangle \nonumber \\
=&\,d^{abc} \fp(y_0) \,,
\label{eq:trBO}
\end{align}
where the second equality implicitly defines $\fp$ (see also eq.~(\ref{fp}) and Appendix~B of Ref.~\cite{deDivitiis:2019xla}).
The left-hand-side consists of correlation functions with one boundary operator and two insertions in the bulk. So the Wick contractions of such a correlation function give:

{\footnotesize{
\begin{align}
a^6 \sum_{\bf x,y} \langle A_0^a(x)  S^b(y) \cO^c \rangle
=&  \dfrac{\mathrm{i} a ^{12}}{L^3} \Tr [T^a T^b T^c] \sum_{\bf x,y} \sum_{\bf u,v} \Bigg \langle \tr \bigg \{ \gamma_0 \gamma_5 [\psi(x) \bar \psi(y)]_{\rm F} [\psi(y) \bar \zeta({\bf u})]_{\rm F} \gamma_5 [\zeta({\bf v}) \bar \psi(x)]_{\rm F} \bigg \} \Bigg \rangle 
\nonumber \\
&  +\dfrac{\mathrm{i} a ^{12}}{L^3} \Tr [T^c T^b T^a] \sum_{\bf x,y} \sum_{\bf u,v} \Bigg \langle \tr \bigg \{ \gamma_0 \gamma_5 [\psi(x) \bar \zeta({\bf u})]_{\rm F} \gamma_5 [\zeta({\bf v}) \bar \psi(y)]_{\rm F} [\psi(y) \bar \psi(x)]_{\rm F} \bigg \} \Bigg \rangle 
\nonumber \\
=& \, \mathrm{i} \Tr [T^a T^b T^c] f_{\rm A S;1} (x_0,y_0)  + \mathrm{i} \Tr [T^c T^b T^a]  f_{\rm A S;2} (x_0,y_0)
\nonumber \\
=&\, \dfrac{1}{2} \bigg [  -d^{abc} \Re f_{\rm A S;1} (x_0,y_0)
 + f^{abc} \Im f_{\rm A S;1} (x_0,y_0) \bigg ] \,.
\label{eq:trBOO}
\end{align}
}}
The second in the above string of equations implicitly defines the two traces of quark propagators (devoid of flavour structure) as $f_{\rm A S;1} (x_0,y_0)$ and $f_{\rm A S;2} (x_0,y_0)$. In the last equation we have made use of the fact that the two traces of propagators are complex conjugates of each other which is a consequence of the 
$\gamma_5$-Hermiticity property of Wilson fermion propagators. Finally, the fact that the above correlation function is invariant under charge conjugation leads to the vanishing of the term proportional to $f^{abc}$ in the last expression. Hence, we obtain
\begin{eqnarray}
a^6 \sum_{\bf x,y} \langle A_0^a(x)  S^b(y) \cO^c \rangle  = - \dfrac{1}{2} d^{abc} \Re f_{\rm A S;1} (x_0,y_0)  =  - d^{abc} \fas (x_0,y_0) \,,
\label{eq:trBOO-2}
\end{eqnarray}
which implicitly defines $\fas (x_0,y_0)$. The correlation functions $\fp$ and $\fas$ are schematically drawn in Fig.~\ref{fig:WI}.
Analogously, from the mass dependent term of the Ward identity we also define $\ftps(x_0,y_0)$; the summation over all times from $t_1$ up to $t_2$ (see eq.~(\ref{eq:SPImassintlatt}))  is included in its definition.
It is important to note that  $d^{abc}$ appears in both eqs.~(\ref{eq:trBO}) and (\ref{eq:trBOO-2}). Therefore, it cancels out in the Ward identity, which becomes
an expression between traces of propagators, without any flavour indices. Putting everything together, we eventually obtain eq.~(\ref{eq:WIf}).

\section{Comparison of $Z$ determinations and scaling tests}
\label{app:z}
\begin{figure}
	\centering
	\includegraphics[width=0.95\textwidth]{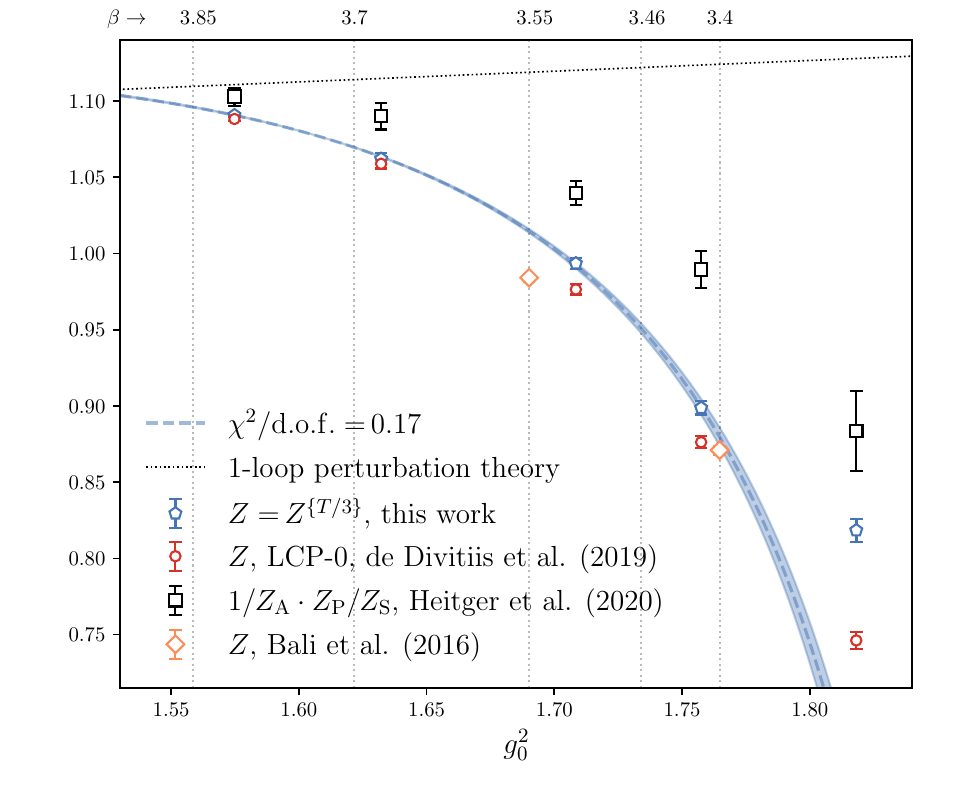}
	\caption{$Z$ results, obtained with different methods, as a function of the squared bare coupling $g_0^2$. 
	The preferred determination of this work is $Z=\zthree$ (pentagons). The squares are obtained by combining results from Refs.~\cite{DallaBrida:2018tpn} and \cite{Heitger:2017njs} (Heitger et al.). The two $Z$ estimates determined in Ref.~\cite{Bali:2016lvx} (Bali et al.) are depicted by triangles. The circles correspond to the $Z$ results from Ref.~\cite{deDivitiis:2019xla} (de Divitiis et al.).  One-loop perturbation theory is illustrated by the dotted line, Ref.~\cite{deDivitiis:2019xla}. The dashed line shows the interpolation (\ref{eq:z_interpolation}) of $\zthree$ (excluding the coarsest lattice spacing from the fit). The vertical dotted lines correspond to the bare couplings used in \textit{CLS} simulations. }
	\label{plot_z}
\end{figure}

In this appendix we present more details on our $Z$ results, listed in Table~\ref{table_am_Z}. In
Fig.~\ref{plot_z} and Table~\ref{tab:Z_comparison} our preferred determination for $Z$, namely $\zthree$ is compared to Ref.~\cite{deDivitiis:2019xla} (de Divitiis et al.),  $Z$ determined at two values of the gauge coupling in Ref.~\cite{Bali:2016lvx} (Bali et al.) and to a $Z$ estimate that we work out from the results of Refs.~\cite{DallaBrida:2018tpn} and \cite{Heitger:2020mkp} (Heitger et al.).
In particular, we extract the axial current normalisation $Z_\mathrm{A}$ at our couplings from  the
interpolation formula of Ref.~\cite{DallaBrida:2018tpn} and combine it with the ratio $Z_\mathrm{S}/Z_\mathrm{P}$  of the pseudoscalar and scalar renormalisation constants from Ref.~\cite{Heitger:2020mkp}. In addition, we give an interpolation formula for our preferred determination for $Z$ (also displayed in Fig.~\ref{plot_z}).

Our result agrees with the other determinations at weaker bare couplings, while disagreements are seen at stronger couplings. These are attributed to lattice artefacts associated with intrinsic ambiguities of $\rmO(a^2)$ or higher between different determinations. Agreement is generally better between our results and those of Ref.~\cite{deDivitiis:2019xla} (de Divitiis et al.).

\begin{table}
	\centering
	\begin{tabular}{lcccc}
		\toprule 
		$\beta$ & \thead{ $Z=\zthree$ \\ this work} & \thead{$Z$, LCP-0 \\ de Divitiis et al.} & \thead{$Z$, LCP-1 \\ de Divitiis et al.} & \thead{$1/Z_\mathrm{A}\cdot Z_\mathrm{P}/Z_\mathrm{S}$ \\ Heitger et al.}  \\ \midrule
		3.3 & 0.8184(77) & 0.7462(56) & 0.7896(36) & 0.884(26)\phantom{0}\\
		3.414 & 0.8987(43) & 0.8762(40) & 0.8992(26) & 0.990(12)\phantom{0}\\
		3.512 & 0.9935(38) & 0.9764(33) & 0.9861(23) & 1.0396(80)\\
		3.676 & 1.0621(36) & 1.0588(31) & 1.0611(23) & 1.0901(89)\\
		3.81 & 1.0907(13) & 1.0882(11) & 1.0884(8)\phantom{0} & 1.1029(61)\\
		\bottomrule
	\end{tabular}
	\caption{Comparison of our preferred $Z$ determination with results from Ref.~\cite{deDivitiis:2019xla} (de Divitiis et al.) and the combination of results from Refs.~\cite{DallaBrida:2018tpn} and \cite{Heitger:2017njs} (Heitger et al.).}
	\label{tab:Z_comparison}
\end{table}

\begin{figure}
	\centering
	\begin{subfigure}[b]{0.49\textwidth}
		\centering
		\includegraphics[width=1.0\textwidth]{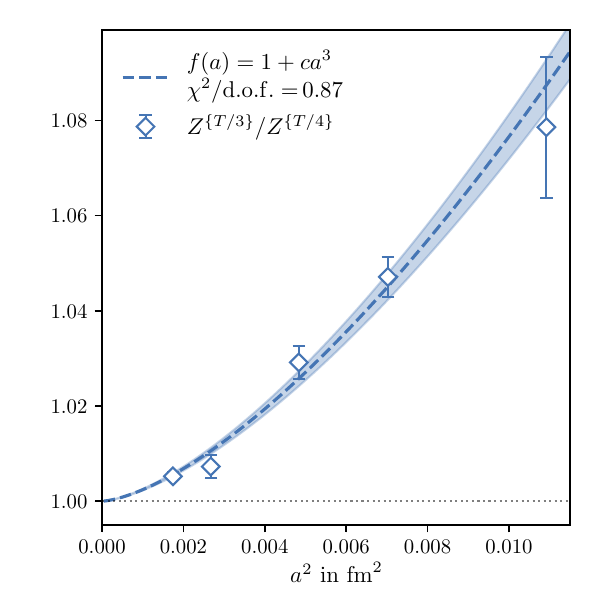}
	\end{subfigure}
	\begin{subfigure}[b]{0.49\textwidth}
		\centering
		\includegraphics[width=1.0\textwidth]{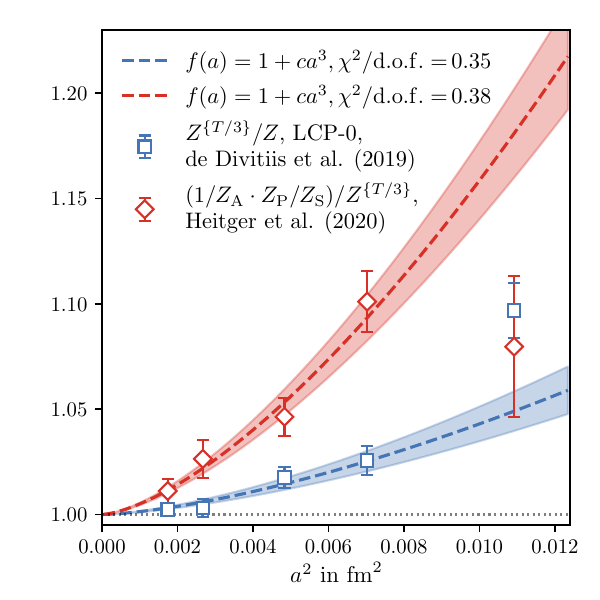}
	\end{subfigure}
	\caption{\textit{Left:} Ratio of our results $\zthree/\zfour$, fitted as a function of the lattice spacing. \textit{Right:} Ratios of $\zthree$ and $Z$ computed in previous works, fitted as a function of the lattice spacing. The coarsest lattice spacing is excluded from the fits. The dashed lines are the fits while the horizontal dotted lines indicate the expected continuum results.}
	\label{z_scaling}
\end{figure}

In order to confirm this claim of consistency (leaving aside higher cut-off effects) we construct ratios of different determinations and investigate their behaviour as a function of the lattice spacing.
Interestingly, rather than $\rmO(a^2)$, leading cut-off effects of $\rmO(a^3)$ can be identified in the ratio $\zthree/\zfour$, as seen in Fig.~\ref{z_scaling} (left). 
The scaling behaviour of our results compared to those of previous works is shown in Fig.~\ref{z_scaling} (right).
All ratios are fitted with an ansatz $1+ca^3$, excluding the coarsest lattice spacing. When adding a term linear in the lattice spacing, its fit parameter vanishes within its uncertainty in all cases. In conclusion, these scaling tests indicate that our results for $Z$ are in accordance with the theoretical expectation of $\rmO(a^2)$ ambiguities or higher which by virtue of the imposed line of constant physics decrease monotically towards the continuum limit.

%\begin{figure}
%	\centering
%	\includegraphics[width=0.95\textwidth]{figures/Z_interpolated}
%	\caption{interpolation of $Z$,interpolation formula~\ref{eq:z_interpolation},the fit parameters~\ref{eq:z_fit_parameters} and the covariance~\ref{eq:z_cov} belong to the red fit that excludes the coarsest lattice spacing, part of figure 8}
%	\label{plot_z_interpolated}
%\end{figure}
In addition, we interpolate our $Z$ data using a Pad\'e ansatz, constrained to the one-loop prediction of Ref.~\cite{Aoki:1998ar} for small couplings; see eq.~(\ref{eq:z_fit}) and Fig.~\ref{plot_z}. Owing to relevant higher order cut-off effects, shown in the right panel of Fig.~\ref{z_scaling}, we do not include the coarsest lattice spacing in the fit. Other than that, it must be noted that the coarsest lattice spacing is well outside the range of CLS couplings. We obtain

\begin{subequations}
	\label{eq:z_fit}
	\begin{equation}
	Z(g_0^2) = 1 + 0.0703169 \cdot g_0^2 \times \frac{1 + d_1 g_0^4}{1 + d_2 g_0^2}\,,
	\label{eq:z_interpolation}
	\end{equation}
	where
	\begin{equation}
	\quad d_i = \left(-0.34504, -0.52309\right)\,,
	\label{eq:z_fit_parameters}
	\end{equation}
	and
	\begin{equation}
	\mathrm{cov}(d_i,d_j) = \left(\begin{array}{lll}
	\phantom{-}2.798505 \times 10^{-7}&\phantom{-}1.054037 \times 10^{-7}\\\phantom{-}1.054037 \times 10^{-7}&\phantom{-}1.545940 \times 10^{-6} % corrected on arxiv in v3
	\end{array}\right).
	\label{eq:z_cov}
	\end{equation}
\end{subequations}
Our $Z$ results at the \textit{CLS} couplings are gathered in Table~\ref{tab:z_cls} and compared to those of Ref.~\cite{deDivitiis:2019xla} for two different LCP conditions. The two outmost \textit{CLS} $\beta$ values ($3.4$ and $3.85$) lie outside the range of our fitted $Z$ estimates, so they are obtained by extrapolation. Their systematic errors are estimated from the statistical uncertainty of the nearest $\zthree$ data point: the systematic error of $\zthree(\beta= 3.4)$ is the statistical error of $\zthree(\beta = 3.414)$ and that of $\zthree(\beta= 3.85)$ is the statistical error of $\zthree(\beta = 3.81)$. These systematic errors are added to the statistical ones in quadrature.

\begin{table}[h]
	\centering
	\renewcommand{\arraystretch}{1.4}
	\setlength{\tabcolsep}{3pt}
	\begin{tabular}{llllll}
\toprule
 $\beta$   & \thead{ $Z=\zthree$ \\ interpolated, this work}   & \thead{$Z$, LCP-0 \\ de Divitiis et al.}   & \thead{$Z$, LCP-1 \\ de Divitiis et al.}   & $ \partial \ln Z /\partial g_0^2$   & $ \partial \ln r_\mathrm{m} /\partial g_0^2$   \\
\bottomrule
 3.4       & $\phantom{-}0.8798(47)(43)[64]$                   & $\phantom{-}0.8758(52)$                    & $\phantom{-}0.8981(35)$                    & $-3.241(144)$                       & $\phantom{-}8.975(195)$                        \\
 3.46      & $\phantom{-}0.9507(25)$                           & $\phantom{-}0.9320(50)$                    & $\phantom{-}0.9468(35)$                    & $-1.974(62)$                        & $\phantom{-}5.915(179)$                        \\
 3.55      & $\phantom{-}1.0147(15)$                           & $\phantom{-}0.9937(42)$                    & $\phantom{-}1.0015(30)$                    & $-1.104(23)$                        & $\phantom{-}3.647(149)$                        \\
 3.7       & $\phantom{-}1.0696(13)$                           & $\phantom{-}1.0591(23)$                    & $\phantom{-}1.0612(17)$                    & $-0.522(7)$                         & $\phantom{-}1.962(90)$                         \\
 3.85      & $\phantom{-}1.0961(12)(13)[18]$                   & $\phantom{-}1.0975(25)$                    & $\phantom{-}1.0971(18)$                    & $-0.278(3)$                         & $\phantom{-}1.190(49)$                         \\
\bottomrule
\end{tabular}

	\caption{
	$Z$ values at the couplings used in \textit{CLS} simulations, obtained from the interpolation formula~(\ref{eq:z_interpolation}), excluding the coarsest lattice spacing from the fit. As explained in the text, an additional systematic error is added to the $\beta=3.85$ and $\beta=3.4$ results and the errors are displayed as: $(\sigma_\mathrm{stat})(\sigma_\mathrm{syst})[\sigma_\mathrm{total}]$. In the last two columns we list $\partial \ln Z /\partial g_0^2$ for $\zthree$, obtained by differentiating eq.~(\ref{eq:z_interpolation}) as well as $\partial \ln \rmsea /\partial g_0^2$ via differentiating eq.~(\ref{eq:fit}).}
	\label{tab:z_cls}
\end{table}
The interested reader may also use our interpolation formula for $Z$ (from eq.~(\ref{eq:z_fit})) and $\rmsea$ (from eq.~(\ref{eq:fit})) and the covariance between the fit parameters of the two different interpolations,
\begin{equation}
\mathrm{cov}(d_i,c_j) = \left(\begin{array}{lll}
\phantom{-}1.810312 \times 10^{-5}&-9.089552 \times 10^{-6}&\phantom{-}5.439349 \times 10^{-9}\\\phantom{-}2.040917 \times 10^{-5}&-1.179257 \times 10^{-5}&-6.592921 \times 10^{-9}
\end{array}\right),
\end{equation}
to construct combinations of the two such as $Zr_\mathrm{m}$.

\end{appendix}
\small
\addcontentsline{toc}{section}{References}
\bibliographystyle{JHEP}
\bibliography{bib}

\end{document}